# MRI Image Generation Based on Text Prompts

Fan Xinxian[1]，Lyu Mengye [1,*]

**1**College of Health Science and Environmental Engineering, Shenzhen

Technology University, Shenzhen, China

*Corresponding author: lvmengye@sztu.edu.cn

**Disclaimer**

This manuscript has been machine-translated from its original Chinese version and may contain inaccuracies.

# Table of Contents

# 摘要

【摘要】 磁共振成像（MRI）是疾病诊断和治疗规划的重要技术，但因采集成本高、罕见病例样本少、患者隐私保护要求高以及依赖专业标识等问题，获取高质量 MRI 影像数据集困难重重，阻碍医学人工智能（AI）研究发展。现有研究大多实现的是无条件自由生成 MRI 图像，这种生成方式缺乏特定的约束和引导，生成图像的应用场景相对受限。鉴于稳定扩散（Stable Diffusion, SD）模型在图像生成领域表现出色，本课题专注于 MRI 图像，研究基于文本提示的 MRI 图像生成算法，根据文本描述生成 MRI 图像，为医学研究提供合成数据集，潜在应用场景更加灵活。通过构建文本与 MRI 图像配对数据集，并采用 DreamBooth 微调以及微调 UNet 两种不同策略微调预训练模型，实现 MRI 图像的可控生成。

本研究采用 FID 和 MS-SSIM 两个定量指标评估生成模型性能。以原始 SD 模型为对照（IncepV3-FID=317.35, CLIP-FID=39.94, MS-SSIM=0.05），使用 DreamBooth 微调后，IncepV3-FID 降至 246.45、CLIP-FID 降至 17.63，但 MS-SSIM 提升至 0.17，表明生成质量提升同时多样性有所下降。研究不同学习率与训练步长组合发现，1000 步短训练步长下，学习率为 $1\times10^{-4}$ 的 FID 和 MS-SSIM 指标更低，模型性能更优，训练步长增至 10000 时，学习率为 $5\times10^{-5}$ 的模型在质量与多样性平衡上表现更好。实验显示需根据具体指标需求调整学习率与训练步长组合，总体而言，微调后模型较原模型在生成质量和语义一致性方面有明显改善。最后，为了说明微调模型在实际下游任务中的实用性，开展 MRI 图像分类实验，实验结果表明使用微调模型生成的图像可以提高图像分类的准确性，在训练样本稀缺时，MRI 图像对比度分类准确率可从 71.15%提高至 96.15%。

【关键词】磁共振图像；图像生成；扩散模型；文本引导

# Abstract

【**Abstract**】 Magnetic resonance imaging (MRI) is an important technology for disease diagnosis and treatment planning. However, due to high acquisition costs, limited rare case samples, high requirements for patient privacy protection, and reliance on professional identification, obtaining high-quality MRI image datasets is difficult, hindering the development of medical artificial intelligence (AI) research. Most existing research has achieved unconditional free generation of MRI images, which lacks specific constraints and guidance, and the application scenarios for image generation are relatively limited. Given the outstanding performance of the Stable Diffusion (SD) model in the field of image generation, this project focuses on MRI images and investigates an MRI image generation algorithm based on text prompts. Based on text descriptions, MRI images are generated to provide a synthetic dataset for medical research and has more flexible potential application scenarios. By constructing a paired dataset of text and MRI images, and using two different strategies of DreamBooth fine-tuning and fine-tuning UNet to fine tune the pre-trained model, controllable generation of MRI images is achieved.

This study used two quantitative metrics, FID and MS-SSIM, to evaluate the performance of the generative model. Compared with the original SD (IncepV3-FID=317.35, CLIP-FID=39.94, MS-SSIM=0.05), after fine-tuning with DreamBooth, IncepV3-FID decreased to 246.45 and CLIP-FID decreased to 17.63, but MS-SSIM increased to 0.17, indicating an improvement in generation quality and a decrease in diversity. Studying different combinations of learning rates and training step sizes, it was found that under a short training step size of 1000 steps, the FID and MS-SSIM metrics with a learning rate of $1 \times 10^{-4}$ were lower, and the model performance was better. When the training step size was increased to 10000, the model with a learning rate of $5 \times 10^{-5}$ performed better in balancing quality and diversity. The experiment shows that the combination of learning rate and training step size needs to be adjusted according to specific indicator requirements. Overall, the fine-tuned model has significant improvements in generation quality and semantic consistency compared to the original model. Finally, to demonstrate the practicality of the fine-tuned model in practical downstream tasks, MRI image classification experiments were conducted. The experimental results show that using images generated by a fine-tuned model can enhance the accuracy of image classification. When training samples are scarce, the accuracy of MRI image contrast classification can be increased from 71.15% to 96.15%.

【**Key words**】 Magnetic Resonance Images；Image Generation；Diffusion Model；Text Guidance

# 1. Introduction

## 1.1 Background and Significance of the Research Topic

In the field of medical imaging, Magnetic Resonance Imaging (MRI) technology [1] is an important tool for clinical diagnosis and medical research, allowing doctors to assess conditions, diagnose diseases, and develop treatment plans. However, due to the high cost of MRI examinations, image acquisition is expensive, which is one reason for the scarcity of MRI image data. Additionally, the scan data for certain rare diseases are very limited, making large-scale collection difficult. Further, strict privacy regulations limit the use of medical data, and manual annotation costs are high, all these factors together make the acquisition of such data extremely challenging [2-3]. Currently, AI-based medical imaging research sometimes faces a data bottleneck, requiring ways to generate diverse and compliant MRI images, so that models have sufficient data for training.

In recent years, generative models such as Generative Adversarial Networks (GANs) [4] and diffusion models [5] have achieved breakthrough progress. The core feature of generative models lies in their unique training process, which enables the generated sample distribution to be as close as possible to the original data distribution. These models have promoted developments in image generation and brought substantial achievements to the field, as well as created possibilities for innovative applications in medical imaging. With the rapid progress of AI technology in medical images, conditional generative models are gradually becoming key technologies to alleviate the insufficiency of medical image data [6-8]. These models generate medical images that meet requirements by providing specific conditions, not only reducing data acquisition and annotation costs, but also providing richer data resources for medical imaging tasks. Among them, text-guided generative models [9] show tremendous application potential: they can effectively utilize semantic information, and the flexibility of textual description ensures high consistency between generated images and text.

Inspired by these achievements, this project studies controllable MRI image generation algorithms, using existing text information to generate high-quality and information-rich MRI images, thus expanding existing MRI image datasets. This algorithm not only efficiently generates image data, but can also generate MRI images with specific modality features under different magnetic field strengths. By using medical imaging terminology and related information as conditional input, the model ensures the medical accuracy of generated images, helps the model understand key image information, promotes the generation of structurally more precise medical images, and provides diverse training materials for medical image processing and diagnostic algorithms, thereby holding considerable significance for medical image data augmentation and other applications.

## 1.2 Current Research Status at Home and Abroad

Text-to-image generation is an important technology in the field of AI. With the rapid development of deep learning and natural language processing, both the requirements and difficulty of text-to-image tasks are continually increasing. The technology can generate images automatically according to natural language descriptions and can be applied to different areas such as artistic generation and computer-aided design. In addition, this technology has promoted multi-modal learning and reasoning research and has become one of the most active research frontiers in recent years [10]. To enable computers to understand and interpret the world from multiple perspectives like humans, multi-modal machine learning was introduced, allowing the modeling, processing, and association of information from different modalities, such as text, images, etc. Text-to-image generation is exactly an embodiment of this concept: it uses AI to concretize text descriptions into images, with the main challenge being to ensure both image quality and the correspondence between the input text and the output image [11].

### 1.2.1 Text-Prompted Natural Image Generation

Benefiting from rapid advances in deep learning, image processing technology and computer vision applications have made significant progress in recent years. Since the proposal of the GAN architecture in 2014 [4] and the emergence of the Deep Dream system in 2015 [12], research on text-guided image generation using deep learning has made major advances [13]. In January 2021, OpenAI released the CLIP model [14], which greatly promoted text-to-image technology and achieved major technological breakthroughs. CLIP is a contrastive language–image pre-training model that classifies images by providing the names of the visual categories to be recognized. In text-to-image contexts, CLIP has been applied significantly for the first time in GAN-based image generation systems [13].

As soon as the CLIP model was released, some AI enthusiasts began building systems that combine GANs and CLIP, specifically for digital art creation. For example, Ryan Murdock's "Big Sleep" attracted large audiences by cleverly combining BigGAN and CLIP [15]. This innovation inspired Katherine Crowson, who further combined the more powerful neural network VQGAN with CLIP [16]. The combination of VQGAN and CLIP quickly became one of the mainstream technologies for generating artistic works, until later surpassed by diffusion models [17]. Nowadays, researchers have richer choices in text-to-image generation systems, but the current frontiers focus on diffusion models, which have attracted wide attention in AI and machine learning. Models such as CLIP-guided diffusion and latent diffusion [18] are leading the way.

## 1.2.2 Text-Prompted Medical Image Generation

Text-to-image technology for natural and medical images has distinct characteristics but also shares some commonalities. In the natural image field, this technology integrates natural language processing and computer vision, and its advances offer technical references to the medical imaging field. High quality and realism are common goals in both fields, both of which need to overcome technical challenges of improving image resolution and detail, and ensuring consistency between generated images and text descriptions is also a shared challenge.

Text-guided medical image generation consists of key steps such as encoding and decoding, including conditional control and parameterization. Compared to natural images, however, medical images are more specialized and structurally more complex, and have more stringent technical requirements. Researchers must keep exploring new technologies and methods to improve image resolution and detail, ensuring high consistency between generated images and textual descriptions.

Currently, conditional generation can be categorized according to the type of conditions used: one category is based on semantic feature constraints, such as organ and tumor maps, which can generate images under specific local constraints [19];

another uses semantically rich and descriptive textual prompts, such as medical text prompts, to enhance the diversity of generated images [20]. For example, in the work of Chambon et al., they used a pre-trained Latent Diffusion Model (LDM), with chest X-ray (CXR) images and corresponding radiology reports as datasets, to generate CXR images of different disease states [20]. Sagers et al. conducted similar studies, using DALL-E to synthesize all Fitzpatrick skin types of skin lesions [21]. To address the problem of class imbalance and reduce the need for manual annotation in medical data, Cho Joseph et al. leveraged the new dataset generation capabilities of LDM to propose MediSyn, a text-guided latent diffusion model for broad 2D and 3D medical modality synthesis [22].

CXR-IRGen [23] and CT-CLIP [24] also demonstrate the latest advances in medical image generation. CXR-IRGen is a model for generating chest X-ray image-report pairs, using a modular architecture with visual and language modules, capable of flexibly generating multi-modal CXR image-report pairs or images/reports alone, and, by combining text and image embeddings, proposes novel prompt design to improve both the overall quality and clinical utility of generated images. CT-CLIP, aiming to provide richer medical image information, integrates 3D medical imaging with text reports to enhance the understanding and processing of medical data, building a model specifically for 3D medical imaging datasets and AI frameworks, in order to improve the accuracy and efficiency of medical diagnoses.

In addition, research has shown that diffusion models can be used for MRI image reconstruction and generation tasks. In MRI reconstruction, Tang et al. combined the latent representation of VAE and diffusion models, using Stable Diffusion (SD) for MRI image reconstruction [25], showing that the SD model has applications on MRI images. Luo et al. introduced a framework in MRI image reconstruction, connecting the diffusion process and generative models with Markov chains to efficiently sample MRI images from learned probabilistic distributions [26], where the process of generating MRI images using diffusion models is also involved. As shown in Figure 1-1, diffusion models can generate realistic MRI images.

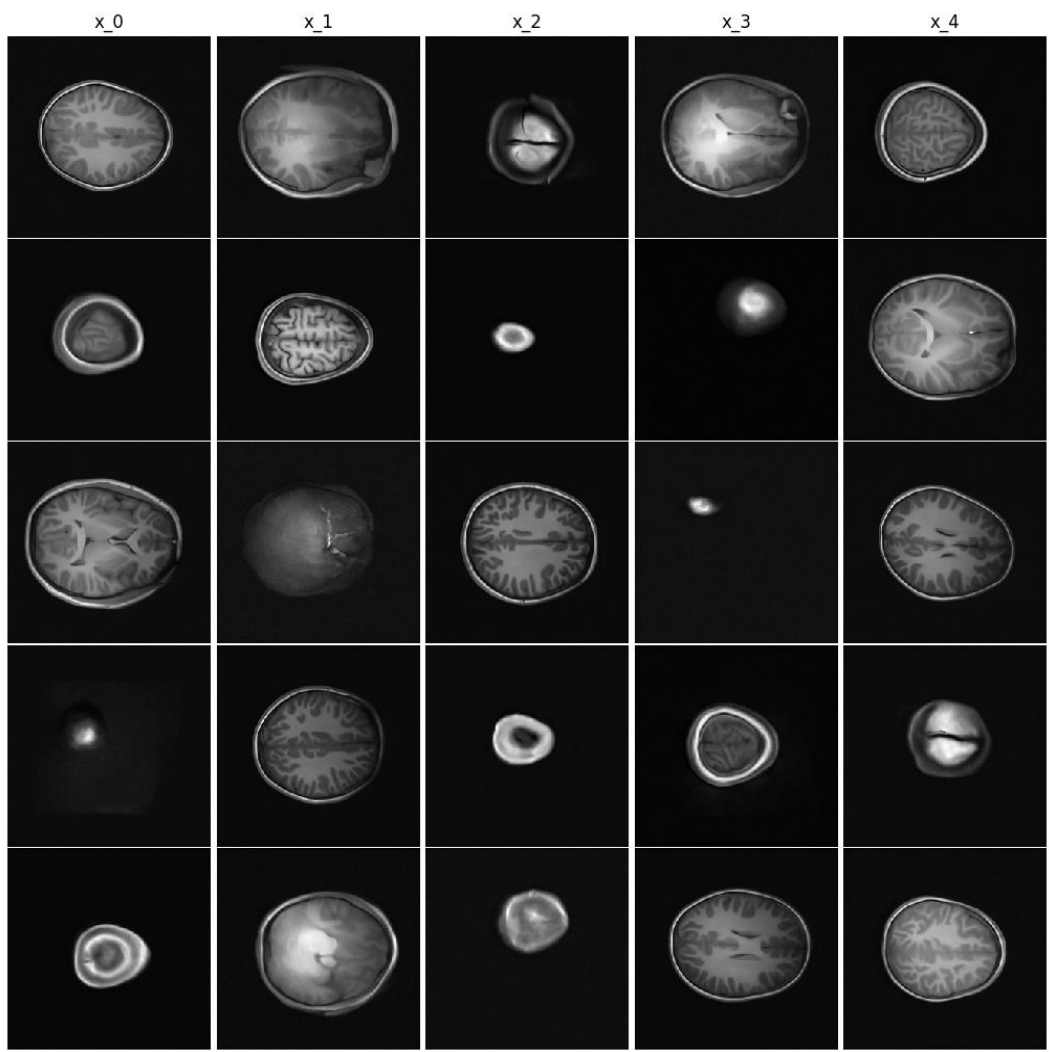

Figure 1-1 Example of MRI image generation using diffusion models from existing research [26]. The drawback is the randomness of image morphology and the inability to exercise fine-grained semantic control.

However, these studies did not make good use of textual information to control the generation of MRI images of different field strengths and modalities. In Tang et al.'s research, the text condition for the SD model was set as null, indicating that no text prompt was used for image generation guidance, thus fine-grained control through text was not possible; Luo et al.'s research used image-based direct diffusion, unable to incorporate text for control, so although the generated MRI images are realistic, they can only be generated randomly or only with data consistency guidance, whereas the field strength, modality, and slice location information cannot be utilized.

# 1.3 Main Research Work and Structure of This Paper

## 1.3.1 Main Research Work

This paper focuses on text-prompted MRI image generation algorithms, covering dataset construction, model selection and training, and generated image evaluation as the three core aspects. By constructing a dataset of paired MRI images of different field strengths and modalities with their corresponding medical imaging descriptions, the model is provided with data for training and evaluation. MRI images are processed into formats suitable for the model, and concise and accurate medical texts are prepared. The Stable Diffusion model serves as the experimental foundation; two fine-tuning methods are used: one using DreamBooth and the other optimizing UNet, to ensure the generated images are visually similar to real images and semantically consistent with the input medical text descriptions. Subjective evaluation is used for qualitative assessment of generated images; quantitative evaluation is performed using Fréchet Inception Distance (FID) and Multi-Scale Structural Similarity (MS-SSIM) to comprehensively assess the authenticity and diversity of generated images, evaluating whether the model's performance meets expectations. Finally, an image classification task using ResNet50 as the classifier is conducted, with accuracy, precision, recall, F1 score, and confusion matrix used to verify the practical effectiveness of the fine-tuned model.

## 1.3.2 Structure of This Paper

This paper consists of six chapters, arranged as follows:

1. Chapter 1 provides an in-depth discussion on the research background and significance, noting the existing flaws of current MRI technologies for AI in medical imaging, and proposing that text-prompted models may offer a solution. Then, the progress of previous studies in this field is reviewed, and the research contents are outlined and summarized.

2. The second chapter delves into the core technologies of MRI and the foundational concepts of diffusion models. It first introduces a brief history of MRI and summarizes the imaging principles of MRI images. Then, it presents basic knowledge about diffusion models and SD models, helping to understand the architecture of the SD model, thus further studying and designing algorithms suitable for this research.

3. The third chapter mainly researches and explains the implementation process of an MRI image generation algorithm based on the SD model, introducing the datasets used in the experiment—M4Raw and fastMRI—the data preprocessing procedure, model fine-tuning strategy, and the evaluation metrics for generated images.

4. The fourth chapter primarily describes the experimental parameter settings, then demonstrates the results of different experiments from both qualitative and quantitative perspectives, making comparisons to conclude that the authenticity of the fine-tuned model has improved significantly, although there is a certain decrease in image diversity.

5. The fifth chapter demonstrates that the fine-tuned model plays a practical role in the classification of different modality MRI images. By training the fully connected layer on pre-trained classification models and comparing models trained on different training sets, it draws the conclusion that images generated by the fine-tuned model can improve the accuracy of classification models.

6. The sixth chapter summarizes the achievements of this research, points out areas for improvement, and looks ahead to future research directions.

# 2. Theoretical Foundations of Text-Prompted Magnetic Resonance Image Generation

## 2.1 Fundamentals of Magnetic Resonance Imaging Technology

### 2.1.1 A Brief History of Magnetic Resonance Imaging

The origin of MRI technology stems from fundamental breakthroughs in physics. In 1946, scientists at Stanford and Harvard Universities first observed the phenomenon of nuclear magnetic resonance (NMR). For their outstanding contributions to the NMR field, they were awarded the Nobel Prize in Physics in 1952. This discovery revealed the rules of energy absorption and release of atomic nuclei in a magnetic field, laying an important foundation for subsequent technological applications. In the 1970s, NMR technology started to permeate the medical field. In 1971, a milestone study [27] detected the relaxation time differences between tumor tissue and normal tissue via NMR, suggesting that this feature could be used for disease diagnosis, marking the nascent stage of medical applications of magnetic resonance technology. Then, in 1973, two researchers independently developed NMR image reconstruction based on gradient magnetic fields [28-29] ; their core theories became the cornerstone of modern MRI technology, earning both the Nobel Prize in Physiology or Medicine in 2003.

With improvements in principles, MRI technology entered a phase of rapid development. In 1977, the world's first whole-body scanner achieved a breakthrough by successfully capturing tomographic images of the human body, turning theory into clinical reality. In the 1980s, devices continually advanced; in 1980, scientists produced the first brain MRI images, and by 1988, breakthroughs in superconducting magnet technology led to the successful development of the first compact MRI device. At this point, MRI had completed its transformation from a laboratory phenomenon to a revolutionary medical tool. The multiple Nobel Prizes awarded throughout its history further highlight the profound influence of this technology on modern medicine.

### 2.1.2 Principles of Magnetic Resonance Imaging

MRI technology mainly utilizes signals produced by water molecules in the body [30]; by using magnetic fields and radio waves to scan the human body, it reconstructs images from signals generated by nuclear resonance in a magnetic field. When operating, a strong magnetic field aligns the orientation of atoms in water molecules, and then the equipment emits specific electromagnetic waves. These atoms absorb energy and emit signals, with the MRI device recording the changes in these signals. The computer system converts them into clear images, displaying the internal structures of the human body.

When the main magnetic field is activated, hydrogen nuclei, due to their intrinsic spin, generate a tiny magnetic moment and start to process around the direction of the magnetic field, gradually aligning with the field. The equipment then emits a specific electromagnetic wave to excite the nuclei within the magnetic field, causing some nuclei to move into a higher energy state. When the electromagnetic wave is turned off, the nuclei gradually return to their original state, emitting detectable radio signals in the process. Signal decay happens in two ways: longitudinal relaxation (T1) and transverse relaxation (T2), which correspond to the recovery of magnetization in the direction of the magnetic field and the decay of the transverse component, respectively.

The MRI system uses gradient magnetic fields that vary along three directions for spatial localization, allowing each signal to correspond to a specific body location. The receiver coil captures these signals, which are digitized and stored in K-space. K-space is a matrix containing the raw MRI image data, with each point carrying spatial and phase information about pixels in the final image. Once the K-space data is fully filled, a Fourier transform converts the different signal components into an image, ultimately generating images that reflect the structure and function of internal tissues.

## 2.2 Overview of Diffusion Models

Diffusion Models [5] have achieved rapid progress in the field of medical imaging in recent years. Their working principle is to gradually add noise to a clear image, making it blurrier step by step until it degenerates into pure random noise; then, a deep neural network is trained to learn the reverse operation, gradually removing noise and finally restoring the image to its original clarity. This process—moving from order to disorder, and then from disorder back to order—provides an innovative approach for high-quality image generation.

Denoising Diffusion Probabilistic Models (DDPM, ( [31] , as one of the pioneering algorithms in the field of diffusion models, consists of two parameterized Markov chains: a forward diffusion process and a reverse denoising process. The forward process involves gradually adding Gaussian noise to the original image until it is completely covered by noise, thereby progressively damaging the original image. In the reverse denoising process, the model learns from the Gaussian noise added during the forward process, training a neural network to recover original data by reversing the noising process. The overall process starts with noise and restores an image with clear visual structure through multiple reverse steps.

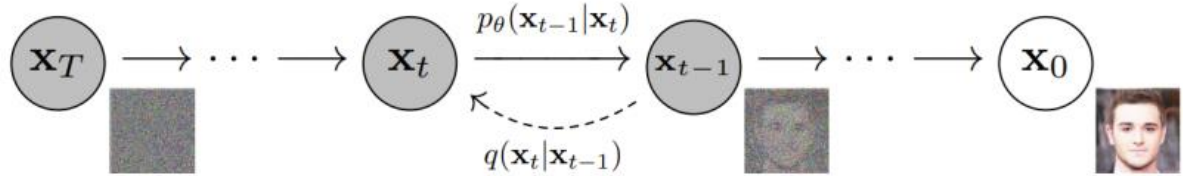

Figure 2-1 Image Generation Process of DDPM [31]

However, DDPM is a relatively inefficient algorithm, so many methods have emerged to improve its generation efficiency. For example, DDIM [32] proposes a more efficient sampling strategy to avoid computing all denoising steps. Additionally, since diffusion models perform Markov-chain-like iterations directly in the image pixel space, they require significant computational resources during training and inference. To address this bottleneck, researchers proposed LDM [18], which performs diffusion steps in latent space, improving image generation while reducing computational cost.

## 2.3 Architecture of the Stable Diffusion Model

SD is an image generation technique based on LDM [18] , which starts with random noise and gradually converts it into the target image to achieve image generation. With the inclusion of conditional mechanisms, the model can control the generation process during inference, for example, by guiding image generation with text prompts and enabling flexible text-to-image translation. Existing pretrained models give the SD model strong generalization capabilities, allowing it to generate specific images based on textual descriptions. In this research, we aim to fine-tune a pretrained SD model, leveraging a curated dataset of MRI image-text pairs to optimize its generation capabilities and achieve MRI image generation.

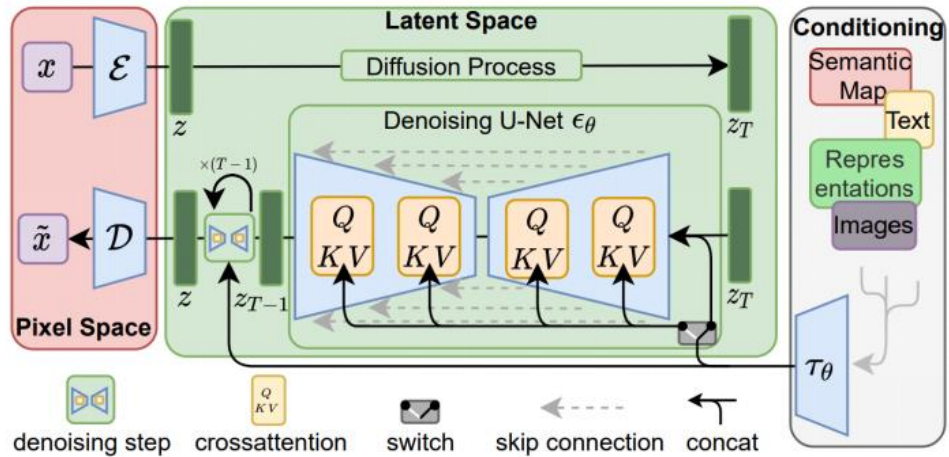

Figure 2-2 Network Architecture Diagram of the Stable Diffusion Model [18]

The SD model image generation workflow is divided into three stages. The text encoder first converts the input text prompt into a high-dimensional semantic vector, which acts as a conditional signal during subsequent generation; the pretrained VAE projects high-dimensional image data into low-dimensional latent space, reducing computational complexity and resource consumption. In latent space, the Markov chain adds Gaussian noise to the latent variables step by step, degrading them into standard normal noise. During reverse denoising, the UNet noise predictor, conditioned on the text semantic vector, iteratively estimates and removes noise, reconstructing a latent representation that matches the text description. During this process, the text condition is injected into multilayer UNet feature maps via a cross-attention mechanism, ensuring precise semantic-image alignment. The denoised latent variables are finally upsampled and decoded by the VAE decoder back to high-resolution pixel space, outputting the target image.

The whole workflow leverages latent space operations and conditional guidance to strike a balance between computational efficiency and controllability, thus ensuring the quality of image generation. This is achieved through the collaboration of three core components, each constructed as a neural network subsystem—known as the three foundational models: the text encoder, the image information creator, and the image decoder.

## 2.3.1 CLIP Text Encoder

CLIP is a multimodal pretrained model [14] comprised of a text encoder and an image encoder. Under the supervision of diverse natural languages, it learns visual features from a wide range of static images and converts textual information into machine-understandable forms. The text encoder adopts a Transformer-based NLP model, responsible for transforming the input text into high-dimensional semantic vectors, while the image encoder uses visual network models like ResNet to extract deep feature representations from images. To achieve semantic consistency between text and images, the model encodes both as vectors to guide the image generation process and understand the relationship between text and images.

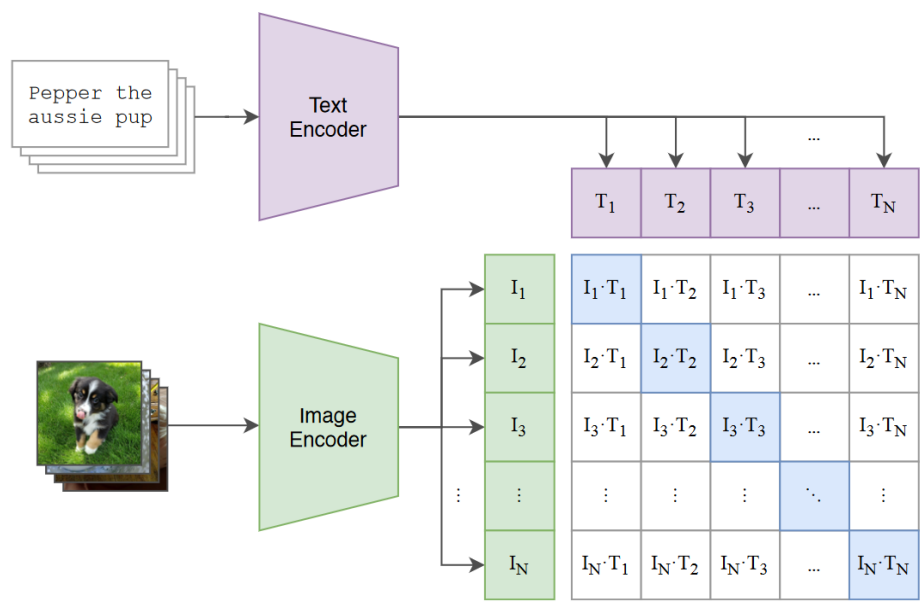

Figure 2-3 Model Architecture of CLIP [14]

The SD model uses a Transformer-based CLIP text encoder, with encoder layers composed of a self-attention mechanism and feed-forward neural network. Input text is converted into tokens by a tokenizer, embedded as vectors, then processed through multiple Transformer encoding layers, outputting a fixed-dimension feature matrix. Through the cross-attention mechanism, this feature matrix interacts dynamically with the UNet visual features; the text semantic vector participates in attention weight calculation as key-value pairs, guiding image detail generation in the denoising process. Thus, natural language instructions are converted into machine-interpretable feature signals to guide the model.

The CLIP model is trained on a dataset containing hundreds of millions of image-text pairs, learning to match image and text semantics, thereby bringing the image features into semantic alignment with their text descriptions and discarding irrelevant feature combinations. After large-scale pretraining, the model achieves strong semantic alignment, so during SD training, the text encoder's parameters are usually frozen to prevent performance degradation and to stably transmit text conditional information.

## 2.3.2 Image Information Creator

The image information creator consists of a UNet and scheduling algorithm. In SD, the UNet model progressively optimizes the image generation by predicting noise and integrates residual networks and cross-attention modules to effectively fuse text information and image features, thereby controlling image generation. The structure of UNet is based on an encoder-decoder architecture but extends it with time embedding and attention mechanisms, enabling dynamic modulation of temporal and textual semantic information during generation.

During training, UNet predicts the noise residual, and together with the scheduling algorithm, gradually removes random Gaussian noise matrices, converting them into latent image features and ensuring optimal image generation through multiple iterative refinements. The training objective is to minimize the error between the predicted and true noise, optimizing model parameters for precise prediction of the noise added at each step. The residual network module in the UNet architecture enhances feature learning via skip connections, while the cross-attention module injects text into the image generation process, strengthening text-based control over image generation. Furthermore, it supports multi-scale training and arbitrary resolution image generation, enabling the SD model to handle data of varying sizes. During inference, UNet repeatedly predicts noise to denoise and generate high-quality latent features, which are then fed into the VAE decoder to reconstruct visible images. The UNet design improves image generation, making the SD generation process more flexible and efficient; once image information processing is complete, results are passed to the image decoder to generate the final image.

## 2.3.3 Image Decoder

At its core, the image decoder is a VAE model, which generates images based on information passed from the image information creator. Like UNet, VAE is designed on the classic encoder-decoder structure and is capable of image compression and reconstruction, enabling mutual conversion between high-dimensional pixel space and low-dimensional latent feature space. In practice, VAE first encodes input images, with the resulting latent features serving as UNet input, and during generation, the processed latent features are restored into complete images.

The VAE architecture includes several innovative components, mainly the GSC module, downsampling module, and upsampling module. The GSC module consists of group normalization, the SiLU activation function, and convolutional layers. The downsampling module uses convolution operations, and during feature decoding, the model restores spatial resolution via an interpolation algorithm combined with convolutional upsampling. To capture global image context, the model integrates a self-attention module at the deep feature processing stage, allowing it to focus on local details as well as overall semantics.

During VAE training, optimization is performed in an encode-decode loop: the encoder maps original images into a low-dimensional latent space, and the decoder transforms latent features back into pixel space. To generate high-quality images and regulate the latent space, the training objective incorporates multiple loss functions for joint optimization: L1 regression loss directly controls reconstruction accuracy, perceptual loss compares feature space representations using a pretrained model for better visual consistency, and the PatchGAN discriminator improves the authenticity of local details from an adversarial learning perspective. The KL regularization term constrains the latent feature distribution, effectively preventing deviation from the normal distribution and providing stable initial conditions for the subsequent diffusion process.

# 3. Design and Implementation of a Text-Prompted MRI Image Generation Algorithm

## 3.1 Dataset Construction and Preprocessing

### 3.1.1 Sources and Selection of Datasets

This study uses two datasets: M4Raw [33] and fastMRI [34] . The M4Raw dataset was acquired using a 0.3T whole-body MRI system from 183 volunteers and contains multi-channel brain k-space data. The acquired image types include T1-weighted, T2-weighted, and Fluid-Attenuated Inversion Recovery (FLAIR) images. This dataset consists of three parts: a training subset, a validation subset, and a motion artifact subset. The training set contains images from 128 volunteers, the validation set contains images from 30 volunteers, and the motion artifact set contains images from 25 volunteers. The fastMRI dataset is a publicly available online dataset with 6970 fully sampled brain MRI data obtained from 3T and 1.5T MRI devices. It contains axial T1-weighted, T2-weighted, and FLAIR images, and some T1-weighted images were acquired with contrast agents.

In this study, the training subset of the M4Raw dataset and the images acquired using 3T MRI devices from the fastMRI dataset (excluding T1-weighted images with contrast agents) were selected. The specific distribution of datasets is shown in Table 3-1.

Table 3-1 Dataset Distribution

| Dataset Partitioning | Image Type | M4Raw Dataset / Images | fastMRI Dataset / Images |
|---|---|---|---|
| Training Set | T1-Weighted Images | 2304 | 2890 |
| | T2-Weighted Images | 2304 | 2890 |
| | FLAIR Images | 2304 | 2890 |
| | Total number | 6912 | 8670 |
| Test set | T1-Weighted Images | 504 | 500 |
| | T2-Weighted Images | 504 | 500 |
| | FLAIR Images | 504 | 500 |
| | Total number | 1512 | 1500 |

To enhance the model's semantic understanding of MRI, this study designs text prompts as an essential component of the multimodal input. Since using DreamBooth to fine-tune the model does not allow each image to correspond to an individual descriptive text, generic concepts are adopted as input text during fine-tuning, which includes information on high and low field strengths as well as different modalities. For the strategy of fine-tuning the UNet, scan parameters are extracted according to the file naming rules and then converted into a text format that the model can comprehend. The key scan parameters are threefold: first is the equipment field strength, with images from the M4Raw dataset coming from 0.3T field strength equipment and the fastMRI dataset belonging to 3T field strength equipment; second is the slice position number; third is the modality type, including T1, T2, and FLAIR modalities.

### 3.1.2 Data Preprocessing

The original experimental data in this study are stored in HDF5 format, a format commonly used for storing scientific datasets and supporting hierarchical metadata management. Since the deep learning model's input interface must be compatible with image formats, this study achieves format conversion through a data preprocessing workflow. According to the modal characteristics of different datasets (M4Raw and fastMRI), the corresponding slice data is extracted using file name identification, and pixel values are mapped to the 0-255 range based on linear normalization, ultimately resulting in a set of standardized two-dimensional grayscale images for subsequent model training. Among them, the image resolution in the M4Raw dataset is 256×256, while in the fastMRI dataset it is 320×320. In later experiments, center cropping is performed on the fastMRI dataset to crop it to 256×256.

In addition, there are issues with some scanning layers in the fastMRI dataset: some slices show blank areas, and others capture only part of the skull. By assessing scan signal strength and structural integrity, it has been found that the first 10 layers of images can display brain tissue structure more clearly. Therefore, a slice retention threshold is set, retaining the first 10 layers for each sample, which effectively filters out unusable scan layers.

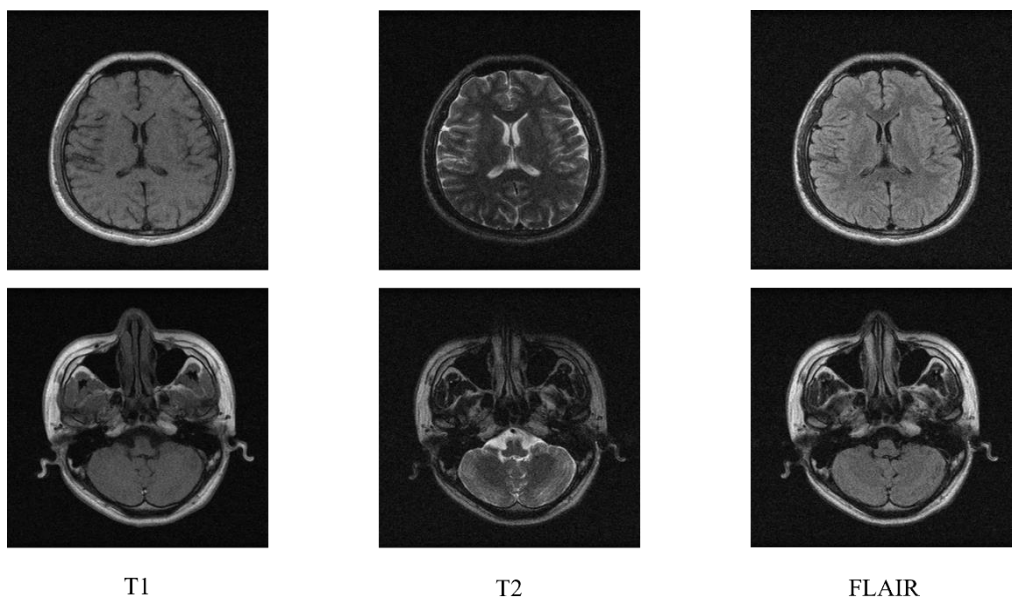

| T1 | T2 | FLAIR |

Figure 3-1 Example of image slices extracted from H5 dataset files

## 3.2 Model Design and Training

### 3.2.1 Model Design and Training Based on DreamBooth Technology

DreamBooth [35] is a full-parameter fine-tuning method for diffusion models, generating high-quality images by embedding theme words and binding rare identifiers. It allows the training process to be completed by updating local parameters on the original SD base model so that the model can learn rare or personalized image features with only a small amount of data, thereby enabling the generated images to add specific themes, objects, or styles more accurately.

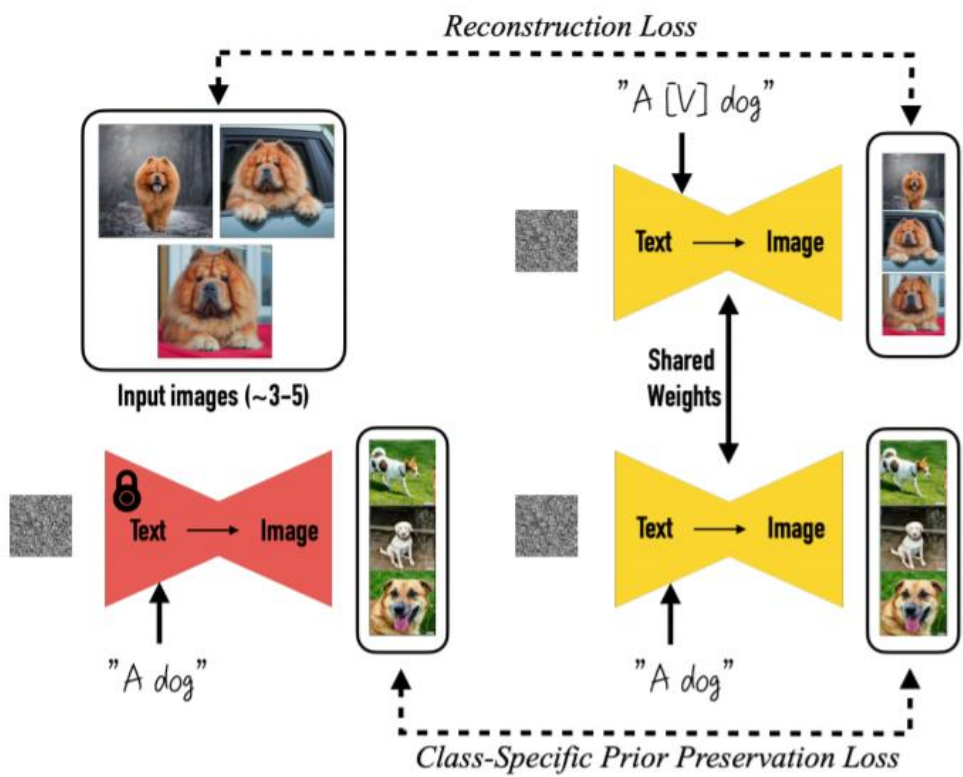

Figure 3-2 Schematic diagram of DreamBooth fine-tuning process [35]

In DreamBooth, all input image text prompts are labeled as "a [identifier] [class noun]", where [identifier] is a unique identifier associated with the subject, and [class noun] is a general category descriptor of the subject. Using class descriptors enables the binding of the category's prior knowledge to the unique identifier of the subject, thus enabling the model to use the visual features of the specific category to generate images consistent with the subject. In addition, DreamBooth found that using existing English words as identifiers did not work well because the model needed to disassociate them from their original meanings and re-associate them with the subject. Therefore, DreamBooth chooses to use rare tokens from the vocabulary as identifiers. This selection of rare identifiers can effectively enable the prior knowledge to be learned, thus improving image generation performance.

There are two major challenges in DreamBooth fine-tuning: language drift and output diversity. Language drift means that the model may forget some of its original abilities during fine-tuning, such as being unable to generate categories other than the target subject. Output diversity refers to a decrease in the variety of model outputs after fine-tuning. In response, DreamBooth adopts self-supervision, using the model's own generated samples to supervise and thereby retain prior knowledge and promote output diversity.

However, in DreamBooth fine-tuning, specific identifiers are primarily associated with target objects through static semantic binding. This technical architecture fine-tunes the UNet and text encoder at the code level and uses text prompt embeddings to intervene in the cross-attention mechanism. This design means that every fine-tuning batch must share the same text prompt, and dynamic text-image pairing is difficult to achieve, resulting in limitations in text condition input dimensionality.

To overcome this limitation, an improved training strategy is adopted here. The dataset is classified according to differences in the physical properties of magnetic resonance imaging, dividing the data into six categories: T1-weighted, T2-weighted, and FLAIR modality images corresponding to 0.3T and 3T devices, respectively. Each subset fully retains original data characteristics without any downsampling. Parameter updates are performed independently for six dedicated models through a unified text guidance strategy. This modality-specific training mechanism not only retains the longitudinal relaxation properties of T1 sequences, the transverse relaxation contrast advantages of T2 sequences, and the cerebrospinal fluid suppression features of FLAIR sequences, but also accurately captures the effect of different magnetic field gradients on proton precession by dividing on the field strength dimension.

## 3.2.2 Model Optimization Based on Fine-Tuning the UNet Module

After preliminary fine-tuning of the SD model with DreamBooth, this study further fine-tunes the core component of the model, UNet. In subsection 2.3.2 above, the functionality of UNet as a key part of the generative model was briefly introduced; in this section, the UNet will be analyzed in more depth from an architectural perspective.

The UNet in the SD model is a two-dimensional, conditionally constrained UNet designed specifically for noise sample conditional generation and reconstruction in SD tasks. The model takes noise samples, time steps, and condition embeddings as input, and outputs denoised features with the same dimensions as the input. The model is structured into three modules: downsampling, middle, and upsampling. The downsampling module abstracts spatial dimensions through hierarchical feature extraction; the middle module fuses cross-modal information; and the upsampling module restores spatial resolution via decoding mechanisms.

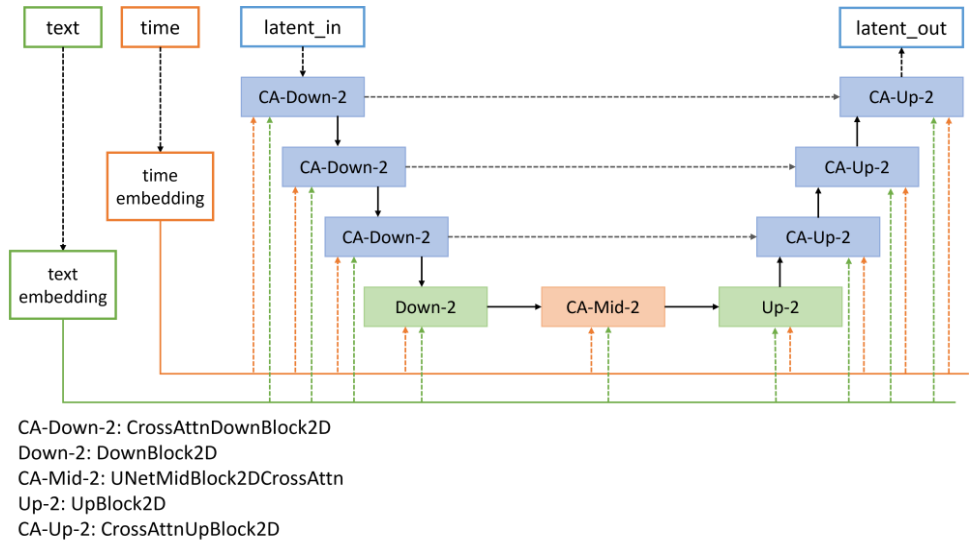

Figure 3-3 Schematic of UNet model architecture

In the downsampling stage, the model uses an improved ResnetBlock2D module to extract features. This module consists of group normalization, SiLU activation function, and convolutional layers. With the help of residual connections, it preserves original features, while the introduction of the Transformer2DModel module normalizes and extracts input data features. This module uses a cross-attention mechanism to match text embeddings and visual features in real time: image features generate query vectors, text embeddings generate key-value pairs, and attention weights associate text semantics with regions in the image. After the attention mechanism, feature normalization ensures stability and consistency. The downsampling is finally completed by Downsample2D layers, which gradually compress feature map dimensions.

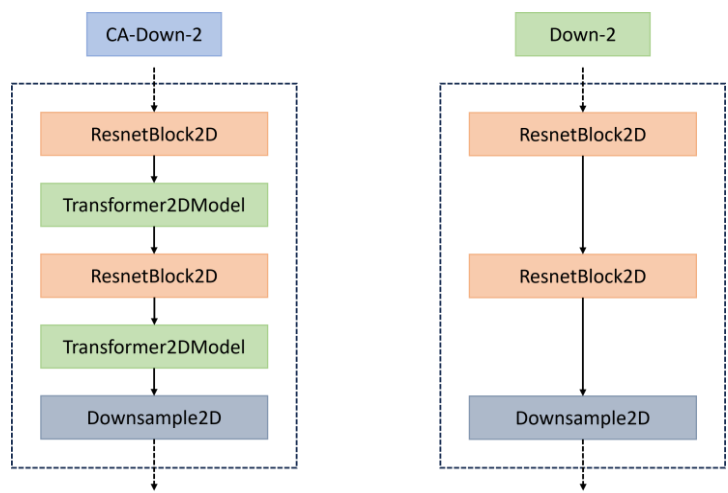

Figure 3-4 Schematic of downsampling module architecture

The middle transition layer connects the encoder and decoder, enhancing global features through concatenated residual blocks and attention modules. The upsampling architecture is similar to the downsampling one. With the help of cross-attention upsampling modules, feature decoding is completed. This module consists of multiple residual blocks and attention modules, alternately extracting and strengthening features to gradually restore details and semantic information during decoding. Skip connections are used between layers to pass intermediate features from downsampling, ensuring that high-frequency details are preserved during reconstruction, thereby improving the clarity and fidelity of the generated images.

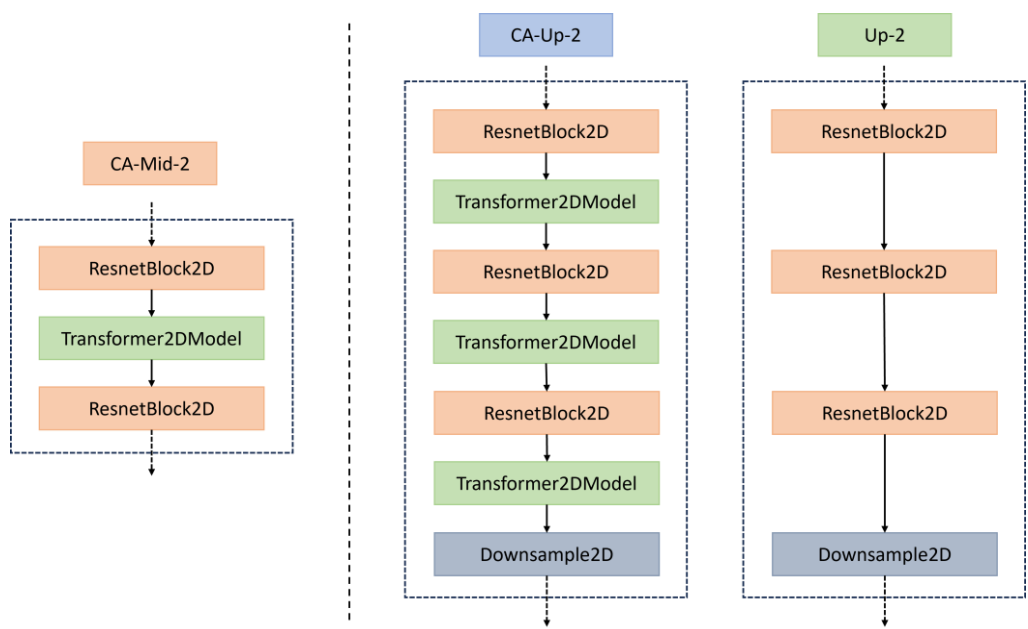

Figure 3-5 Schematic of the middle and upsampling module structure

# 3.3 Evaluation Metrics for Generated Images

## 3.3.1 Fréchet Inception Distance (FID)

FID is a metric used to evaluate the performance of generative models. It measures the difference between the distributions of generated and real data by calculating the Fréchet distance between two distributions. In SD and other diffusion models, FID effectively assesses the similarity between generated images and real images. FID was developed as an improvement to the Inception Score; Martin Heusel et al. [36], when studying GAN performance evaluation, observed some limitations in the Inception Score and therefore proposed FID as a more comprehensive and reliable metric.

The core idea of FID is based on the Fréchet distance: if the distribution of generated image sets is closer to that of real image sets in the feature space, the generated images are more similar to real ones and of higher quality. The feature vectors are extracted from the intermediate layers of the Inception model, hypothesized to be Gaussian distributed, and the means $\mu$ and covariance matrices $\Sigma$ of generated and real data are calculated. Finally, the FID formula is defined as the Fréchet distance between two Gaussian distributions. The calculation formula is as follows:

$$FID = \|\mu - \mu_w\|_2^2 + Tr\left(\Sigma + \Sigma_w - 2(\Sigma\Sigma_w)^{1/2}\right) \qquad (3\text{-}1)$$

Where, $\mu$ and $\mu_w$ are the mean vectors of the generated data and real data, respectively; $\Sigma$ and $\Sigma_w$ are the covariance matrices of the generated data and real data, respectively; $Tr$ denotes the trace of the matrix; $(\Sigma\Sigma_w)^{1/2}$ is the square root of the product of covariance matrices. The smaller the FID value, the closer the distribution of generated data is to the real data, and the better the performance of the generative model. The advantage of FID is that it considers both the diversity of the generated data (covariance matrix) and its similarity to real data (mean difference), and FID is relatively robust to noise and disturbances, accurately reflecting the quality of generated data.

Since FID is typically calculated based on the Inception V3 model pretrained on the ImageNet dataset, it may display certain limitations in domain adaptation scenarios, especially when processing multimodal MRI image data and may not effectively capture the characteristic information of MRI data under different field strengths. Therefore, in this study, the FID score is calculated using the intermediate layers of two completely different models in terms of architecture, domain, and pretraining strategies: the Inception V3 model pretrained on the ImageNet dataset and the CLIP-ViT-B-32 model trained on a large number of text-image pairs.

### 3.3.2 Multi-Scale Structural Similarity (MS-SSIM)

Multi-Scale Structural Similarity (MS-SSIM) is also an advanced metric [37] for image evaluation, designed to overcome the limitations of single-scale structural similarity (SSIM) when dealing with different resolutions and viewing conditions. MS-SSIM is based on the human visual system's (HVS) strong adaptability to structural information and provides a more flexible and human perception-aligned image evaluation method by combining structural similarity measurements across multiple scales. Compared with single-scale SSIM, it offers a more comprehensive and finer evaluation of images, better reflecting variations at different resolutions and the human eye's sensitivity to structural information at different scales.

This evaluation model is designed according to the characteristics of human vision. It separates image properties such as luminance, contrast, and structural features layer by layer and then measures their similarity with the original image to evaluate image quality. Luminance similarity $l_M(x, y)$ is calculated at scale M, assessing the overall lighting consistency of two images by differences in local mean distribution at this scale; contrast similarity $c_j(x, y)$ and structural similarity $s_j(x, y)$ are calculated at scale $j$ for contrast and structural feature similarity, respectively. Contrast similarity is measured by comparing the reference image $x$ and the distorted image $y$ variance at scale $j$, and structural similarity is measured by comparing the covariance of the reference image $x$ and the distorted image $y$ variance at scale $j$ at scale, reflecting the similarity in structure.

The model first applies low-pass filtering and downsampling to the reference and distorted images to progressively extract image features at different scales. At each scale, contrast and structural similarity are computed, and at the final scale, luminance similarity is also computed. The final multi-scale SSIM index is obtained by a weighted combination of measurements at different scales:

$$MS - SSIM(x, y) = [l_M(x, y)]^{\alpha_M} \cdot \prod_{j=1}^{M} [c_j(x, y)]^{\beta_j} [s_j(x, y)]^{\gamma_j} \qquad (3\text{-}2)$$

where $\alpha_M$、 $\beta_j$、 $\gamma_j$ these parameters represent the weights of different scales, reflecting the human visual system's sensitivity to features at different scales.

# 4. Experimental Results and Discussion

## 4.1 Experimental Design and Parameter Settings

This study adopts two fine-tuning strategies to optimize the model. The specific experimental design is as follows: First, a basic fine-tuning framework is built based on DreamBooth, and model parameters are fine-tuned to establish a baseline model. Further, the UNet architecture is fine-tuned with a focus on optimizing parameters of the cross-attention layer and residual connection modules. To explore the effect of training dynamics on model performance, a two-factor controlled experiment is designed in which the learning rate is set as $\alpha \in \{1 \times 10^{-4}, 5 \times 10^{-5}\}$, and the maximum training steps are set as $T_{max} \in \{1000, 10000\}$. By combining different learning rates and maximum training steps, four training configurations (M1－M4) are formed. The experimental group consists of six groups: the original pretrained model, DreamBooth fine-tuning model, and UNet fine-tuning models (M1–M4) to form controlled experiments. During DreamBooth fine-tuning training, a constant value strategy is used for learning rate scheduling to prevent overwriting existing knowledge and to ensure gradual convergence during training, thereby improving training stability and effectiveness. When fine-tuning UNet, a cosine learning rate scheduler is used to dynamically adjust the learning rate during training, which improves model convergence and helps avoid local optima. Other major fine-tuning parameters are shown in Table 4-1.

Table 4-1 Parameter settings for different fine-tuning strategies

| Fine-tuning | Learning Rate | Training Steps | Loss Function | Optimizer | Batch Size | Gradient Accumulation Steps |
|---|---|---|---|---|---|---|
| DreamBooth | $5 \times 10^{-6}$ | 400 | Mean Squared Error (MSE) | AdamW | 1 | 1 |
| UNet | $1 \times 10^{-4}$ | 1000 | Mean Squared Error (MSE) | AdamW | 1 | 4 |
| | $5 \times 10^{-5}$ | 1000 | | | | |
| | $1 \times 10^{-4}$ | 10000 | | | | |
| | $5 \times 10^{-5}$ | 10000 | | | | |

The experiments run on a Linux system with Python 3.9.7 and PyTorch 2.5.1+cu11.8, trained in single-precision floating point on six Tesla P40 GPUs. Distributed training is implemented with the accelerate library (v1.2.1) to enable multi-GPU parallel computation, speeding up training and optimizing model convergence. The SD model weights are obtained from the "stable-diffusion-v1-5/stable-diffusion-v1-5" repository on Hugging Face Hub, and the code is written based on the diffusers library. For DreamBooth fine-tuning, the model training uses a learning rate of $5\times10^{-6}$, the maximum training steps are set to 400, batch size 1, gradient accumulation steps 1, and each model training lasts about 15 minutes. For UNet finetuning, learning rates are set to $1\times10^{-4}$ and $5\times10^{-5}$ with maximum training steps of 1000 and 10000, batch size 1, and gradient accumulation steps up to 4 due to memory usage. Training with 1000 steps takes about 1 hour and 30 minutes, while 10000 steps take about 16 hours and 30 minutes.

## 4.2 Loss Function Curves

Mean Squared Error (MSE) is used as the loss function during fine-tuning, which helps the model generate images that better meet the requirements. The model collects the predicted noise and real noise, calculates the MSE value between them to update model parameters, so that with each generation, the model gets closer to the characteristics of real images, eventually generating realistic images that match text descriptions. Figures 4-1 and 4-2 show the loss function curves for the DreamBooth fine-tuned model and the separately fine-tuned UNet component.

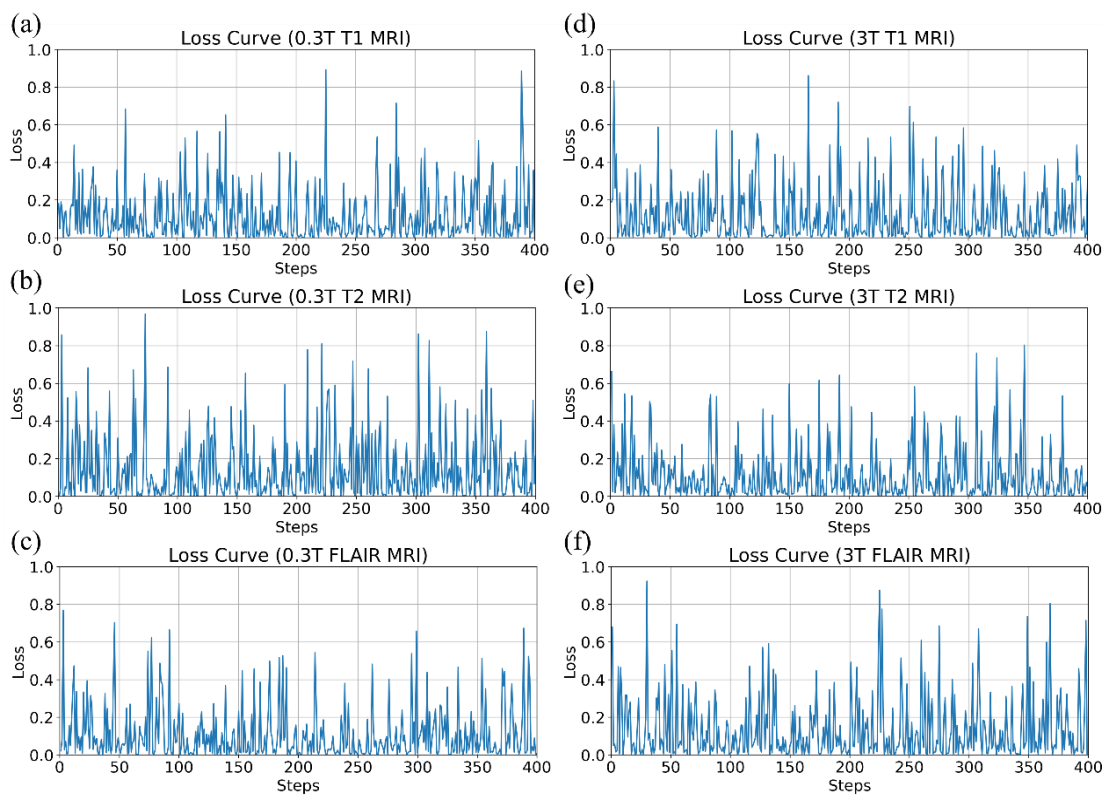

Figure 4-1 Loss function curves of each model after DreamBooth fine-tuning. (a–c) show the loss function curves of the T1, T2, and FLAIR models under low field strength; (d–e) show those under high field strength for the same three modalities.

For the six models fine-tuned using DreamBooth, the loss function curves all show varying degrees of fluctuation. In the T1 and T2 modalities, the amplitude of fluctuation under low field strength is greater than that under high field strength, indicating that under low field strength, image features are relatively blurry and noisier, making stable learning difficult and requiring continual adjustment during training, hence the more volatile loss curves. However, for the FLAIR modality, fluctuation amplitude is greater under high field strength than low, possibly because high field strength images in this modality, while having clearer details, also contain more complex features and contrast variations, thus requiring more adaptation during training and showing larger fluctuations.

Compared to DreamBooth fine-tuning, the loss function curves for fine-tuning the UNet component show significantly reduced fluctuations, with loss values basically below 0.5. All four models show large initial fluctuations that lessen over time, indicating instability in early training as the model learns new features each epoch. As training proceeds and with more steps, the process stabilizes and the models learn features of each modality under different field strengths well. The gap between predicted and real noise decreases, so the fluctuation amplitude also decreases. It is also found that, in the UNet fine-tuning strategy, different learning rates have little influence on the loss curve, likely due to the use of the AdamW optimizer, which adaptively adjusts each parameter's learning rate.

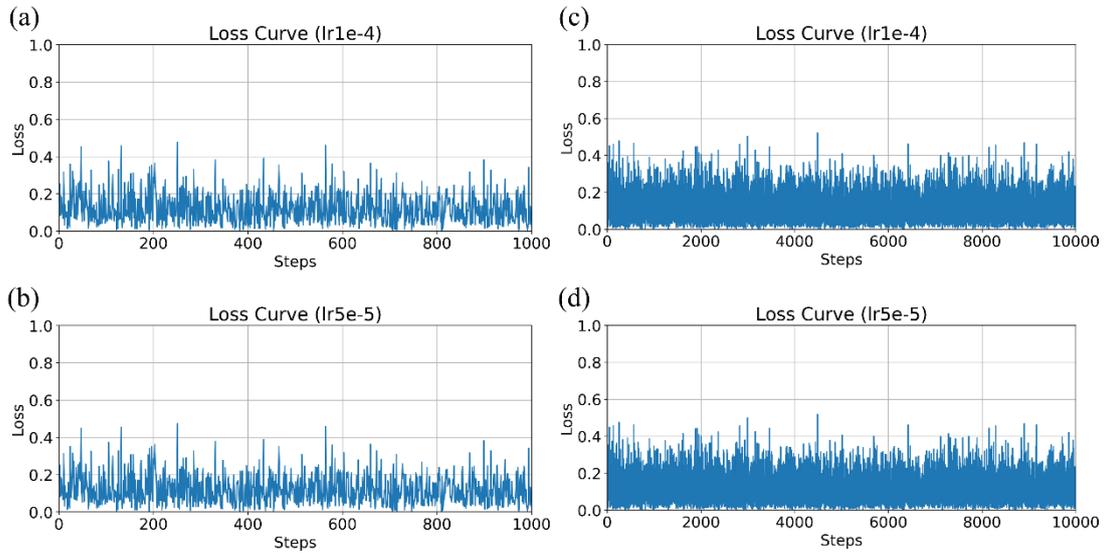

Figure 4-2 Loss function curves after UNet fine-tuning. (a) shows the loss curve for the model with a learning rate of $1 \times 10^{-4}$ and 1000 training steps; (b) for $5 \times 10^{-5}$ and 1000 steps; (c) for $1 \times 10^{-4}$ and 10000 steps; (d) for $5 \times 10^{-5}$ and 10000 steps.

## 4.3 Qualitative Evaluation of Generated Images

This study conducts multimodal MRI image generation experiments under different field strengths using six model architectures to evaluate MRI image generation performance of each. Visual comparison is used to observe the structural accuracy, edge sharpness, and noise control ability (caused by field strength) for each modality generated by different models. This section presents representative generated samples and preliminarily explores the effect of model fine-tuning on the images, providing an intuitive basis for subsequent quantitative analysis.

### 4.3.1 Overview of Generated Images

Figure 4-3 shows a comparison of MRI images generated by the original SD model and the SD model fine-tuned with DreamBooth. The left shows images generated by the original SD model, while the right shows those from the DreamBooth fine-tuned model. It can be seen that although the original SD model can generate brain-related images, this is based on pretraining with natural images, so the model itself does not know the characteristics of different field strengths or modalities. Thus, generated images are brightly colored and structurally unrealistic, whereas after DreamBooth fine-tuning, the images are closer to real MRI images in color, structures are clearer and details more realistic, better matching the imaging characteristics of MRI. This indicates that the DreamBooth fine-tuning strategy can effectively improve the fidelity of generated images, making them closer to the distribution of real data.

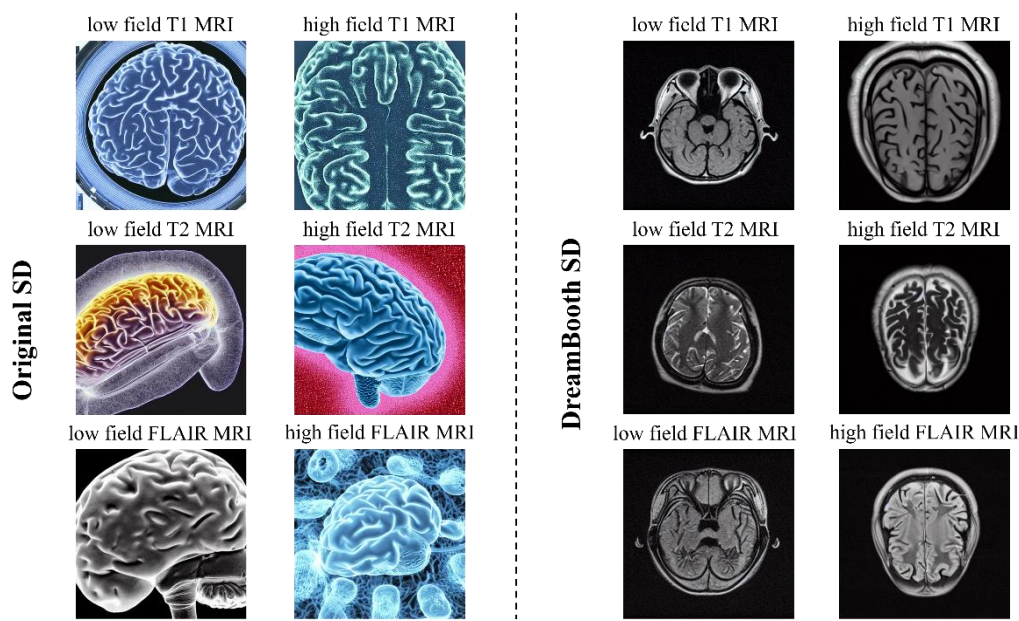

Figure 4-3 Comparison of generated results between the original model and the DreamBooth fine-tuned model

This study uses text-image paired datasets to specifically fine-tune the UNet network, exploring the impact of text conditions on image generation results. By comparing images generated by models fine-tuned with different parameters, the role of textual prompts in image generation is analyzed. As shown in Figure 4-4, overall, the fine-tuned SD model performs well for MRI image generation, effectively expressing the data features of different modalities and generating T1, T2, and FLAIR images highly similar to real ones.

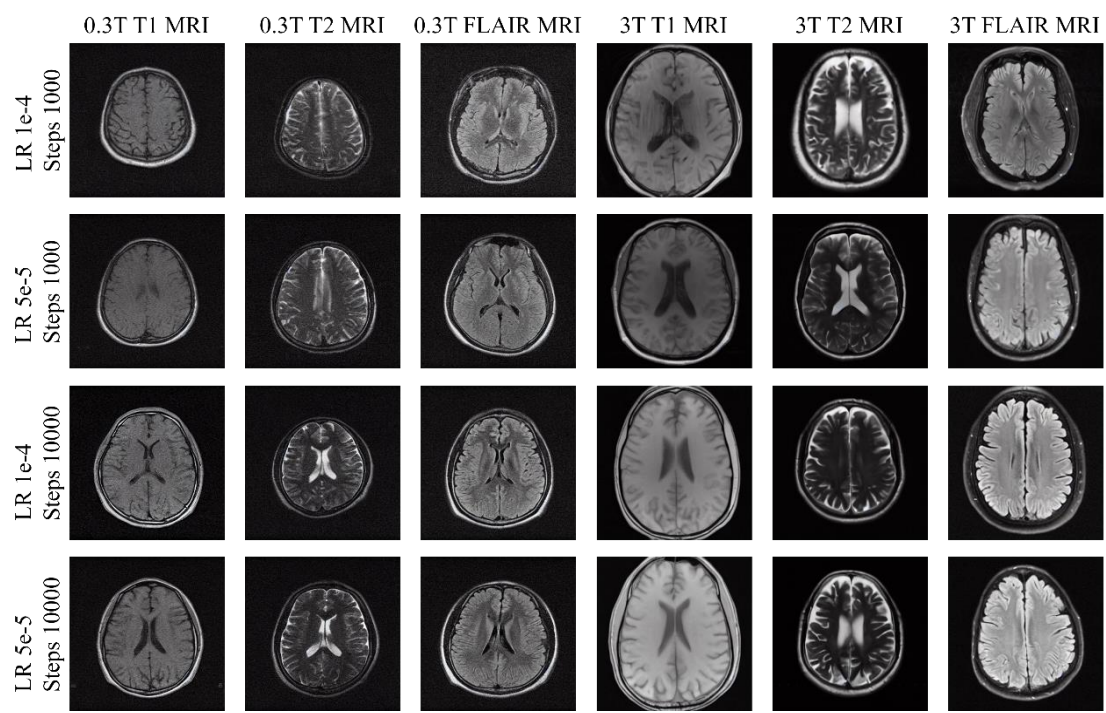

Figure 4-4 Generated image results with different parameters

With a learning rate of $1\times10^{-4}$ and steps of 1000, images of each modality at 0.3T field strength show some noise but retain detail; at 3T, visual differences in detail are observed compared with the other three groups, but in general the imaging characteristics of each MRI modality are preserved. When the learning rate is $5\times10^{-5}$ and the steps are 1000, noise in 0.3T images is reduced and contrast improved; at 3T, head and brain tissue edges are clearer and details are more pronounced. Increasing steps to 10000 provides further improvement. The two groups of images with 10000 steps are sharper in texture than those with 1000 steps, and images generated with the two learning rates under 10000 steps look similar. Compared to DreamBooth fine-tuned results, images generated by fine-tuning the UNet are closer to real images, which also suggests that detailed textual prompts provide some constraint and guidance for SD model image generation. In addition, each model shows good generalization ability, adapting to different field strengths and modalities, and generating images that match the corresponding imaging characteristics, enabling effective generation of specific modality images under diverse conditions.

## 4.3.2 Comparison with Real Images

To further assess the images generated by the model, the plan is to input text information about specific layers and verify whether model-generated images correspond to real images at the same layer. By comparing similarity with real images, we further understand model generative ability under different fine-tuning parameter combinations. It is important to note that differences between individual anatomical structures may exist due to diversity in anatomy or external factors, so even at the same anatomical layer, real images may show different features.

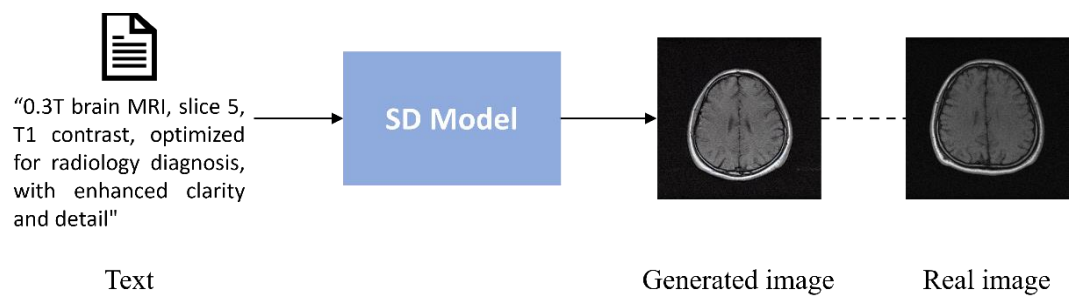

Figure 4-5 Model generation performance verification process

The original model and the DreamBooth fine-tuned model do not involve information about image layers, therefore, only the four UNet fine-tuned models are compared. Figures 4-6 to 4-10 show images generated by the four models for different slices, with one example for each category. Overall, from the generated images, the models show their own characteristics under different modalities and field strengths, and the details and structures of the generated images are highly similar to those of the real images.

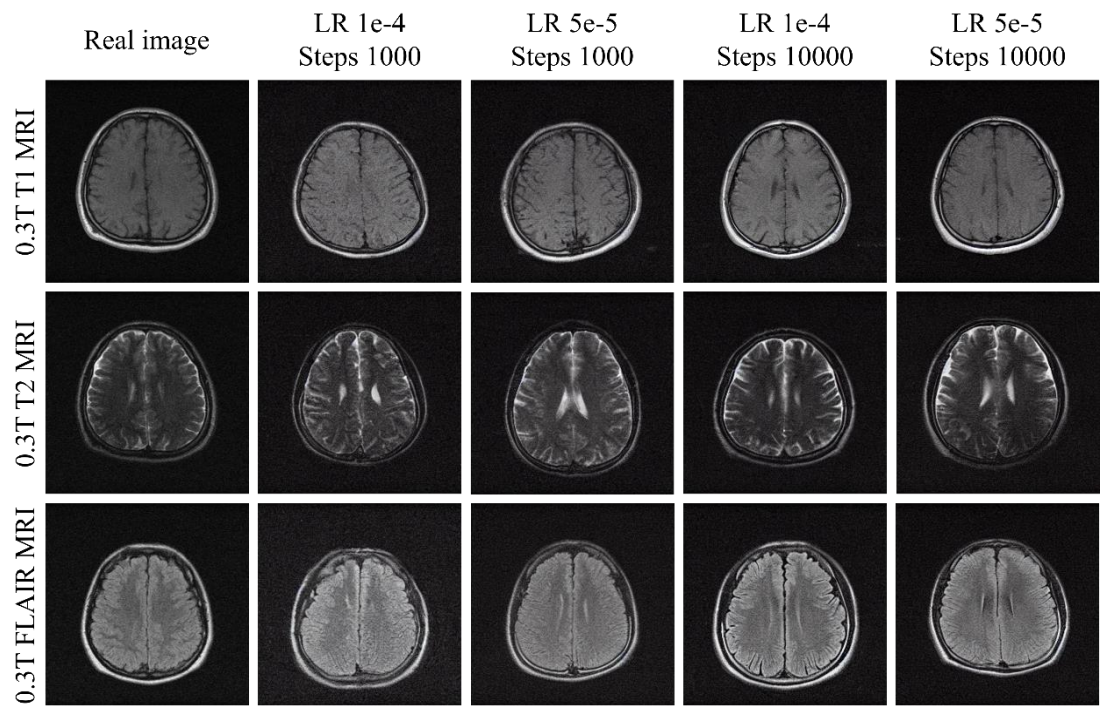

Figure 4-6 Generation results for low field strength images by each model at the same fifth slice

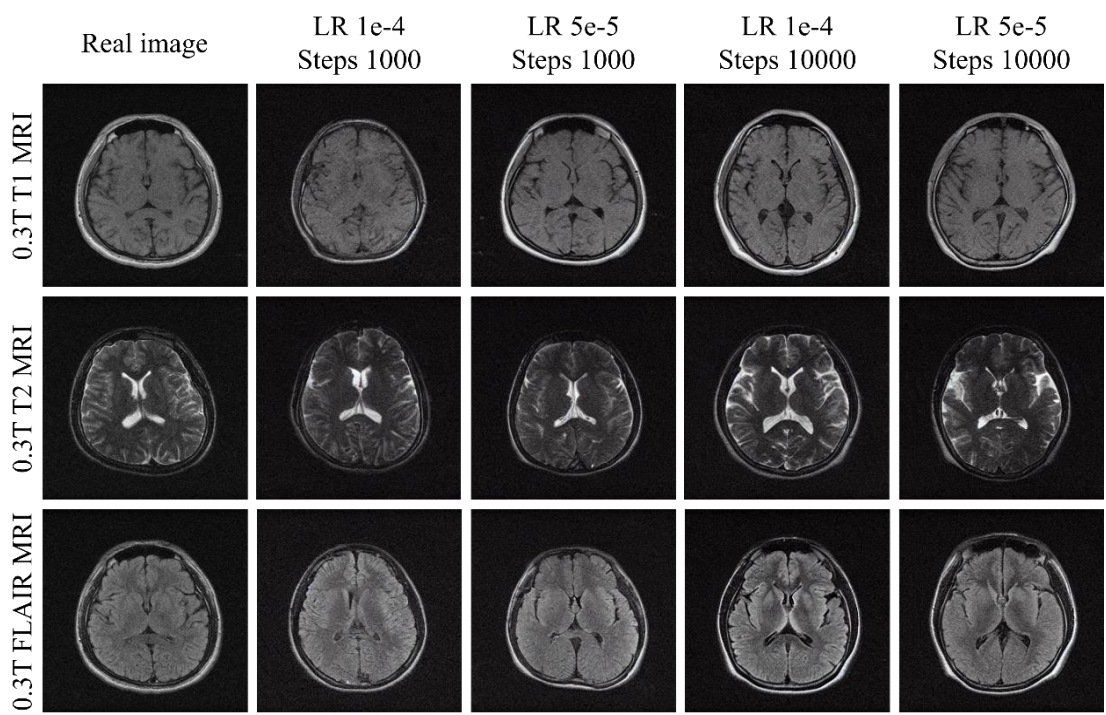

Figure 4-7 Generation results for low field strength (0.3T) images by each model at the same ninth slice

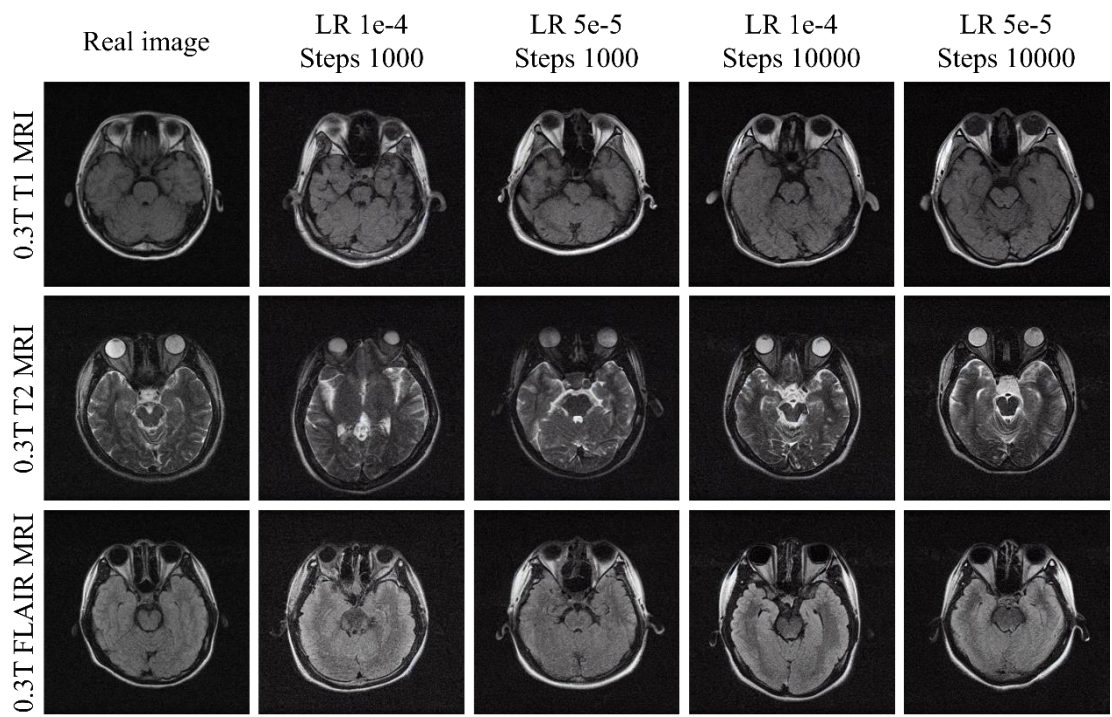

Figure 4-8 Generation results for low field strength (0.3T) images by each model at the same thirteenth slice

Each MRI image in the M4Raw dataset has 18 slices. Here, the fifth, ninth, and thirteenth slices are selected for representative display, as these three slices have different image features, allowing for a more intuitive assessment of whether the model has sufficiently learned features at different layers. The figures clearly show that overall features for each modality are present, but there are also differences from real images, indicating that the model can distinguish characteristics of different modalities and is partly aware of features for each slice. For the three modalities, images generated with a learning rate of $1 \times 10^{-4}$ and steps of 1000 have the most noise. With increased learning rate and steps, noise gradually decreases. For 10000 steps, T2 modality shows higher signal than real images and is more concentrated, while FLAIR modality also has higher contrast than real images. This may be because the model focuses more on local information, leading to generation of higher-frequency details.

In the fastMRI dataset, each MRI image only retains the first ten slices, so the second and fifth slices are displayed here. As in the M4Raw dataset, the model is able to learn the features of different modalities and layers, and the generated images have high similarity with real images. The brightness issue is also present in the T2 and FLAIR modalities of high field strength images. In terms of noise, since the fastMRI dataset images are collected by high field strength devices, the real images are relatively clear, and generated images have almost no noise in the background. However, the brightness of generated images varies under different parameter settings, and the parameters for each modality under different layers are also different, suggesting that parameters have a certain impact on brightness.

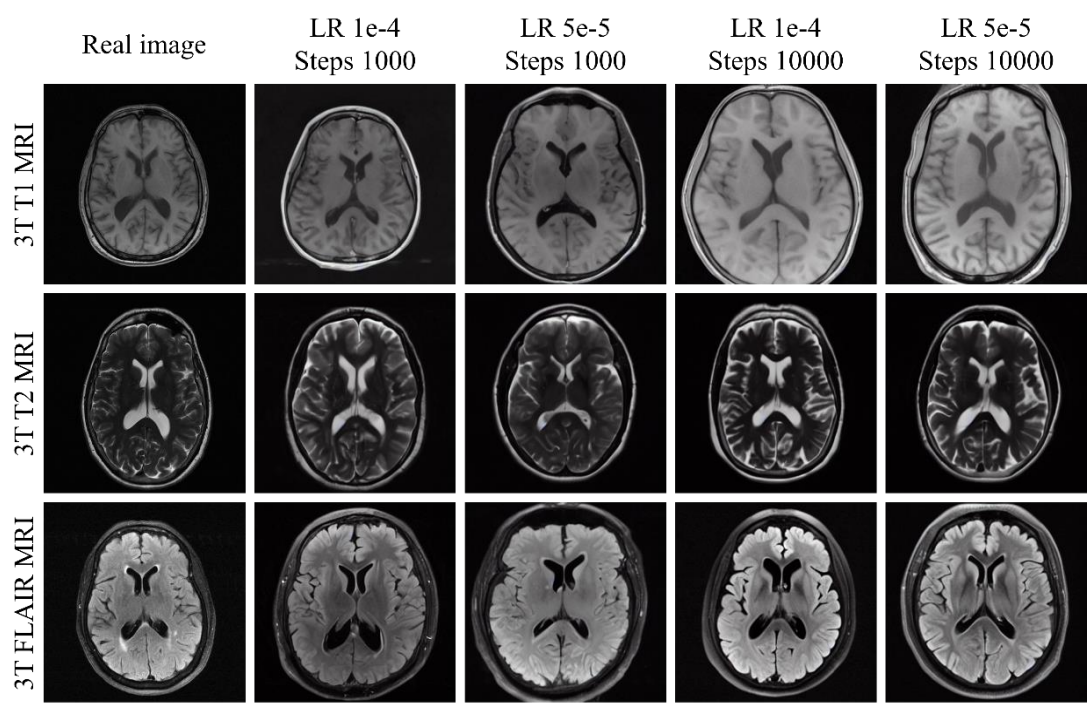

Figure 4-9 Generation results for high field strength (3T) images by each model at the same second slice

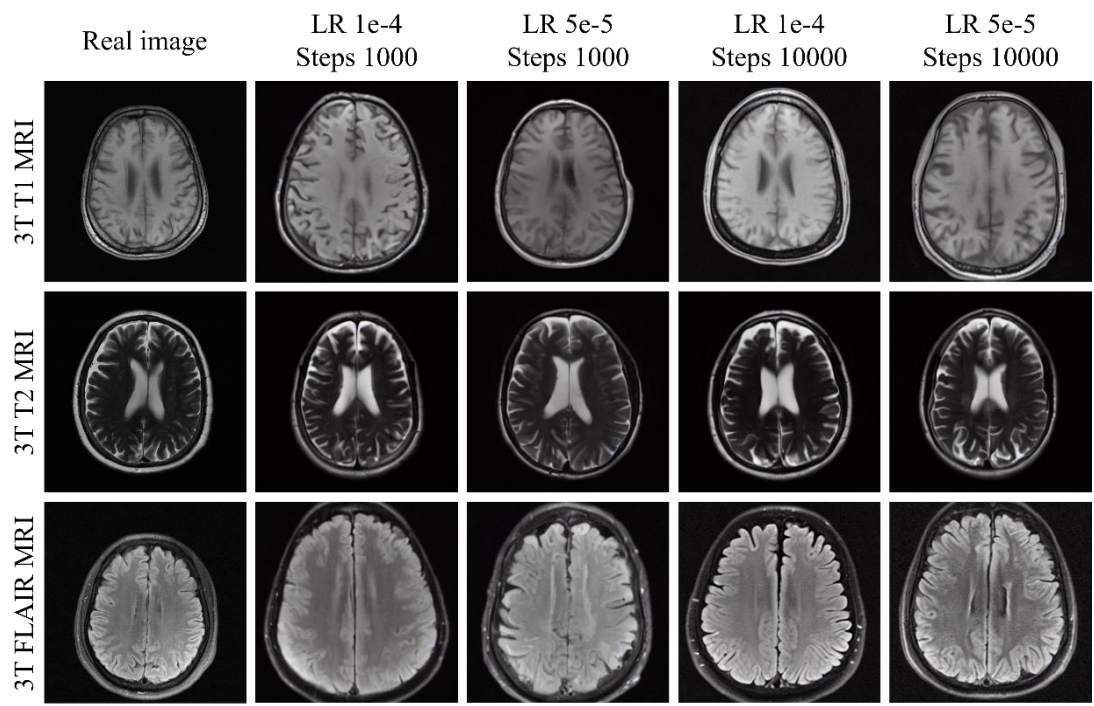

Figure 4-10 Generation results for high field strength (3T) images by each model at the same fifth slice

Overall, the images generated by each model at the same slice have a certain degree of similarity, indicating that the model can accurately identify the features of images at different slices. In terms of brightness, real images have a relatively uniform brightness distribution, clear details, and moderate contrast. In contrast, the brightness of generated images is more affected by learning rate and training steps, and the images generated with 10000 steps are clearer than those with 1000 steps.

### 4.3.3 Effects of Different Prompts on Generated Images

The prompts used during fine-tuning are described in Section 3.1.1. On this basis, the text content was supplemented in two ways: one is by adding imaging requirements to the original prompt, and the other is by including features not present in the prompt during fine-tuning, to observe whether the model can generate the corresponding MRI images.

Figures 4-11 to 4-13 show generation results using (1) the same prompt as in training, (2) prompts with added imaging requirements, and (3) prompts with features not seen in training on the DreamBooth fine-tuned model. From the results, such textual

changes have little impact on the generated images. When the prompt contains features not seen in training, since the model did not learn those features, the images generated are similar to those seen in training. Similar experiments on the UNet fine-tuned models yield the same results.

prompt: "low/high field T1/T2/FLAIR magnetic resonance image"

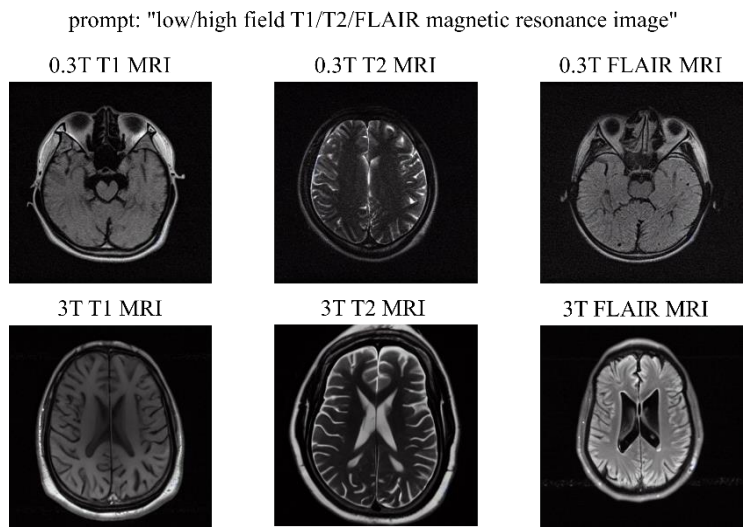

Figure 4-11 Generation result using the same prompt as in training

prompt: "low/high field T1/T2/FLAIR MRI image with clear anatomical details"

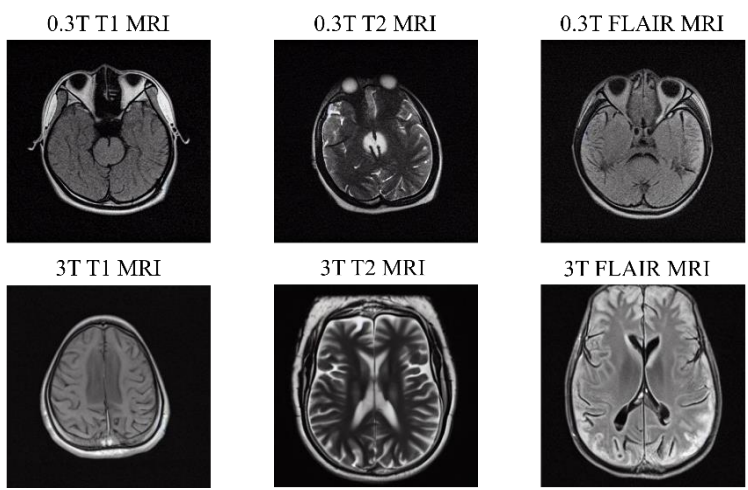

Figure 4-12 Generation result with prompts containing imaging requirements

prompt: "low/high field T1/T2/FLAIR magnetic resonance image with tumor"

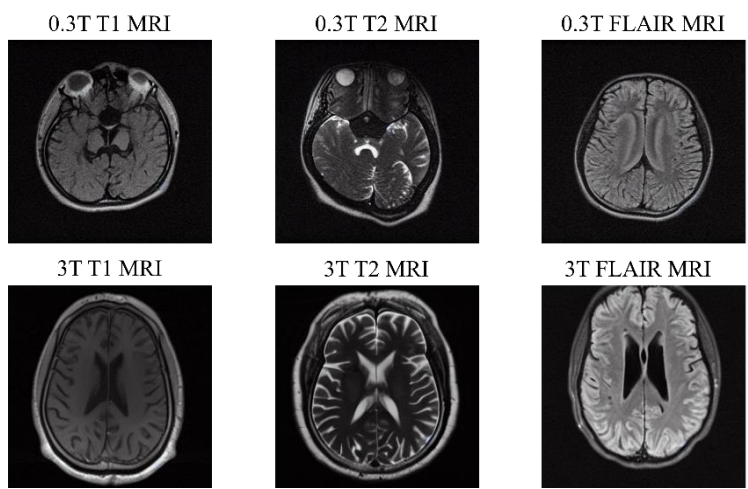

Figure 4-13 Generation result with prompts containing features not seen in training

# 4.4 Quantitative Evaluation of Generated Images

This study quantitatively evaluates the quality and diversity of generated images under different fine-tuning strategies and experimental conditions. To comprehensively measure the similarity between generated and real images, FID was used as one of the evaluation metrics, with lower FID values indicating that the distribution of generated and real images is closer. To evaluate the diversity of generated images, MS-SSIM is also used, which reflects diversity by measuring the structural similarity between generated images—a lower value indicates greater diversity. Table 4-2 presents the evaluation metrics for each group.

Table 4-2 Quantitative evaluation of image generation model fidelity and diversity (FID/MS-SSIM)

| Experiment | FID | | MS-SSIM |
|---|---|---|---|
| | IncepV3 | CLIP | |
| **Baseline** | | | |
| Original SD | 317.35 | 39.94 | 0.05 |
| DreamBooth SD | 246.45 | 17.63 | 0.17 |
| **Learning Rate, Training Steps** | | | |
| 1e-4, 1000 | 88.18 | 9.98 | 0.29 |
| 5e-5, 1000 | 91.64 | 10.72 | 0.38 |
| 1e-4, 10000 | 107.60 | 12.58 | 0.31 |
| 5e-5, 10000 | 102.04 | 11.53 | 0.30 |

As shown in the table data, the FID value of the original SD model is high without fine-tuning, indicating that the generated images deviate significantly from the real data distribution. This reflects that general pretrained models, without specific domain knowledge, struggle to capture task-specific detailed features. The low MS-SSIM value of the original SD model shows weak structural similarity among generated images, possibly due to random generation methods that ensure diversity but lack controllability. After DreamBooth fine-tuning, the IncepV3-FID and CLIP-FID scores of the SD model drop to 246.45 and 17.63 respectively, improving by 22.4% and 55.9% compared to the original SD model, which demonstrates that fine-tuning enables the model to quickly learn key features of the target domain. At the same time, fine-tuning raises the MS-SSIM value, indicating that the model shifts from random to semantically guided generation, enhancing target structural features and making generated images structurally more consistent.

Further fine-tuning of the UNet component can further improve the FID value of DreamBooth SD. With 1000 training steps, the model with a learning rate of $1\times10^{-4}$ outperforms the $5\times10^{-5}$ model in terms of FID, indicating that a relatively high learning rate within a short training period can accelerate the model's convergence to the target domain feature distribution and alleviate the representation bias of the original SD in the specific data domain. When training steps increase to 10000, the FID rises slightly, possibly because the model shifts from fitting a few images to generating more diverse compositions, thus reducing the quality of individual results. Under 10000 training steps, the model with a lower learning rate performs better, implying that a lower learning rate helps alleviate overfitting during training. As shown in the table, all fine-tuned models have significantly higher MS-SSIM values than the original SD model, indicating improved structural similarity in generated images after fine-tuning, but with reduced diversity.

### 4.4.1 FID Indicator

The FID results in Table 4-3 show that different fine-tuning strategies have a significant impact on multimodal MRI image generation, and there is a complex nonlinear relationship between learning rate and training steps.

Table 4-3 FID Scores for Each Modality under Different Models

| Modality | Original SD | DreamBooth SD | 1e-4,1k | 5e-5,1k | 1e-4,10k | 5e-5,10k |
|---|---|---|---|---|---|---|
| **IncepV3** | | | | | | |
| 0.3T_T1 | 339.20 | 296.91 | 87.83 | 103.54 | 115.73 | 100.90 |
| 0.3T_T2 | 313.72 | 210.35 | 94.71 | 111.49 | 121.49 | 124.80 |
| 0.3T_FLAIR | 361.01 | 263.41 | 91.94 | 106.55 | 132.51 | 116.70 |
| 3T_T1 | 303.33 | 304.41 | 106.97 | 83.37 | 113.44 | 116.23 |
| 3T_T2 | 283.18 | 222.52 | 77.55 | 73.66 | 71.92 | 72.93 |
| 3T_FLAIR | 303.63 | 181.12 | 70.09 | 71.23 | 90.52 | 80.66 |
| **CLIP** | | | | | | |
| 0.3T_T1 | 39.09 | 15.68 | 10.67 | 12.00 | 13.49 | 11.50 |
| 0.3T_T2 | 44.03 | 19.19 | 10.09 | 11.46 | 14.11 | 14.13 |
| 0.3T_FLAIR | 40.00 | 16.62 | 9.26 | 10.65 | 14.09 | 12.75 |
| 3T_T1 | 42.41 | 21.70 | 11.15 | 10.93 | 13.14 | 13.02 |
| 3T_T2 | 35.32 | 17.23 | 7.80 | 8.35 | 7.60 | 7.79 |
| 3T_FLAIR | 38.82 | 15.38 | 10.87 | 10.92 | 13.03 | 9.99 |

The IncepV3-FID values of each modality for the original SD model exceed 280, with most CLIP-FID values above 35, indicating that the general pretrained model has image feature distortion problems when generating medical images with different field strengths. After DreamBooth fine-tuning, the model shows improvement in most modalities, for example, the 0.3T T1 images' IncepV3-FID drops from 339.20 to 296.91, and CLIP-FID from 39.09 to 15.68, demonstrating that fine-tuning improves image realism and strengthens the semantic consistency between text prompts and generated images. However, for the 3T field strength T1 modality, fine-tuning led to a decline in IncepV3-FID performance, suggesting that this modality's anatomical complexity or incompatibility with current fine-tuning strategies requires further parameter adjustment.

Further analysis of the synergy between learning rate and training steps reveals that the model with a learning rate of $1\times10^{-4}$ and 1000 steps adapts better to low field strength MRI images, as their IncepV3-FID and CLIP-FID values are the lowest among the four models compared. For high field strength MRI images, the T1 modality images perform best with a learning rate of $5\times10^{-5}$ and 1000 steps, with IncepV3-FID and CLIP-FID values of 83.37 and 10.93, respectively; for T2 modality, the lowest FID values are also with a learning rate of $1 \times 10^{-4}$ and 1000 steps; for FLAIR modality, the lowest IncepV3-FID occurs at a learning rate of $1 \times 10^{-4}$ and 1000 steps, and the lowest CLIP-FID at a learning rate of $5 \times 10^{-5}$ and 10000 steps, indicating that in this modality, a high learning rate may optimize visual features in the short term, while a low learning rate may better optimize semantic consistency in the longer run.

Comparing different training steps, when the training steps are 1000, the model with a learning rate of $1 \times 10^{-4}$ performs better during training than the original and DreamBooth fine-tuned models. For example, in the 0.3T FLAIR modality, IncepV3-FID drops from 361.01 in the original model to 91.94; however, when steps are extended to 10000, some modalities experience overfitting, causing the FID to rise again. Compared to the model with a learning rate of $1 \times 10^{-4}$, the $5 \times 10^{-5}$ model shows better optimization at 10000 steps than at 1000 steps.

## 4.4.2 MS-SSIM Indicator

MS-SSIM evaluates the diversity of generated images by comparing the structural similarity among images in the generation set. The lower the value, the more diverse the generated images, which is significant in assessing medical image generation models.

Table 4-4 MS-SSIM Scores for Each Modality Under Different Models

| Modality | Original SD | DreamBooth SD | 1e-4,1k | 5e-5,1k | 1e-4,10k | 5e-5,10k |
|---|---|---|---|---|---|---|
| 0.3T_T1 | 0.06 | 0.14 | 0.35 | 0.45 | 0.40 | 0.38 |
| 0.3T_T2 | 0.05 | 0.23 | 0.40 | 0.47 | 0.42 | 0.44 |
| 0.3T_FLAIR | 0.05 | 0.28 | 0.36 | 0.45 | 0.39 | 0.36 |
| 3T_T1 | 0.05 | 0.14 | 0.18 | 0.27 | 0.19 | 0.18 |
| 3T_T2 | 0.05 | 0.05 | 0.30 | 0.35 | 0.33 | 0.31 |
| 3T_FLAIR | 0.05 | 0.17 | 0.17 | 0.27 | 0.13 | 0.15 |

Since the original SD model does not have text-image pair constraints, the MS-SSIM value is 0.05, showing a remarkably low value, which suggests the generated images are highly diverse. However, such unconstrained generation can lead to semantic deviations from requirements, reflected in poor FID scores. With the introduction of approximate text constraints through DreamBooth fine-tuning, the MS-SSIM values increased to varying degrees in all except high field strength T2 modality, indicating a certain suppression of diversity, but the improved FID confirms the effectiveness of the constraints by significantly increasing the similarity to anatomical structures.

Due to the more specific text conditioning during UNet fine-tuning, compared to the DreamBooth fine-tuned model, models with different parameter combinations undergo changes in diversity. With a learning rate of $1 \times 10^{-4}$, the MS-SSIM value increases with the training steps; however, for a learning rate of $5 \times 10^{-5}$, the model performs better with longer steps, with MS-SSIM values lower than those with 1000 steps, effectively improving generation diversity in all modalities. Under the same training steps, shorter steps at a learning rate of $1 \times 10^{-4}$ result in better diversity, with each modality's MS-SSIM lower than at $5 \times 10^{-5}$; conversely, with 10000 steps, the $5 \times 10^{-5}$ learning rate yields more diversity in most categories. These results indicate that parameter configuration also affects diversity, and selecting an appropriate combination of learning rate and steps can better meet task requirements.

# 5. Practical Application Showcase and Effectiveness Verification of the Obtained Model

## 5.1 Experiment Objective

For the fine-tuned models, in order to explore their effectiveness in practical applications, an image classification experiment with different modality MRI images was designed as an example. Classification models were trained on different training sets, and their classification results on the test set images were compared with ground truth, to preliminarily assess whether generated images can help improve classification accuracy.

## 5.2 Data Source and Preprocessing

The dataset uses 0.35T MRI image data from Huayin City Hospital in Shaanxi Province, 3T MRI image data from Shenzhen BGI Research Institute, and 0.3T MRI images generated by the fine-tuned model (learning rate $1\times10^{-4}$, training steps 1000). The 0.35T MRI data includes T1, T2, and FLAIR modalities for five individuals, and the 3T data is for one individual's three modalities. The 0.35T data is in DICOM format, and the 3T data in NIFTI format; both need to be converted to PNG and resized to 256×256 to standardize classification model input. After conversion, each 0.35T subject has 13 slices, each 3T subject has 18 slices. Based on dataset sizes, the fine-tuned model generated 13 slices per modality, with prompts like "0.3T brain MRI, slice x, T1/T2/FLAIR contrast", where x is a random number between 1 and 18. Four subjects' data from the 0.35T device were selected as the test set, and four training sets were used for comparative training as shown in Table 5-1: the first set used data from the remaining 0.35T subject, the second set used the 3T dataset, the third set used synthetic 0.3T images generated by the fine-tuned model, and the fourth used a combination of the first and third sets.

Table 5-1 Distribution of Classification Model Datasets

| Dataset Partitioning | Modality | Number / Images |
| --- | --- | --- |
| Training with 0.35T Real Data | T1 | 13 |
| | T2 | 13 |
| | FLAIR | 13 |
| Training with 3T Real Data | T1 | 18 |
| | T2 | 18 |
| | FLAIR | 18 |

Continuation of Table 5-1 Dataset Distribution

| Dataset Partitioning | Modality | Number / Images |
| --- | --- | --- |
| Training with Synthetic Data | T1 | 13 |
| | T2 | 13 |
| | FLAIR | 13 |
| Training with 0.35T Real + Synthetic Data | T1 | 26 |
| | T2 | 26 |
| | FLAIR | 26 |
| Test set | T1 | 52 |
| | T2 | 52 |
| | FLAIR | 52 |

## 5.3 Experiment Setup

The experiments use ResNet50 for classifying different MRI modalities, initializing model weights with ImageNet pretraining. During training, convolutional layers are frozen and only the fully connected layers are optimized, to mitigate overfitting risks in small sample scenarios. The fully connected layer output dimension is set to 3 for T1, T2, and FLAIR, producing four classification models, which are then evaluated on the test data for classification accuracy.

In the data loading phase, considering limited sample size, images are augmented with rotation and scaling. Classification labels are determined by image names: images with "T1" are labeled 0, "T2" as 1, and "FLAIR" as 2. The training set is split 8:2 into training and validation sets to monitor accuracy during training. The learning rate is set to $5 \times 10^{-4}$, training for 100 epochs, using Adam optimizer and cross-entropy loss.

# 5.4 Experiment Results

## 5.4.1 Training Process

As mentioned, the dataset is split into training and validation sets; the model learns features and patterns from the training set, and is then evaluated on the validation set. A confusion matrix is calculated to show the relationship between predictions and real labels. For visualization, heatmaps are used for the confusion matrix, as shown in Figure 5-1.

As seen from the figure, models trained on the four different training sets all perform well on their respective validation sets; all non-diagonal elements of the confusion matrices are zero, indicating accurate predictions for all categories—the model can correctly distinguish T1, T2, and FLAIR modalities in the validation set. Each category can be clearly separated, and no sample from one category is misclassified as another, which demonstrates the model's effective learning of image features from the training data. Notably, the extremely high validation accuracy on small data may indicate overfitting, requiring evaluation on an independent test set later.

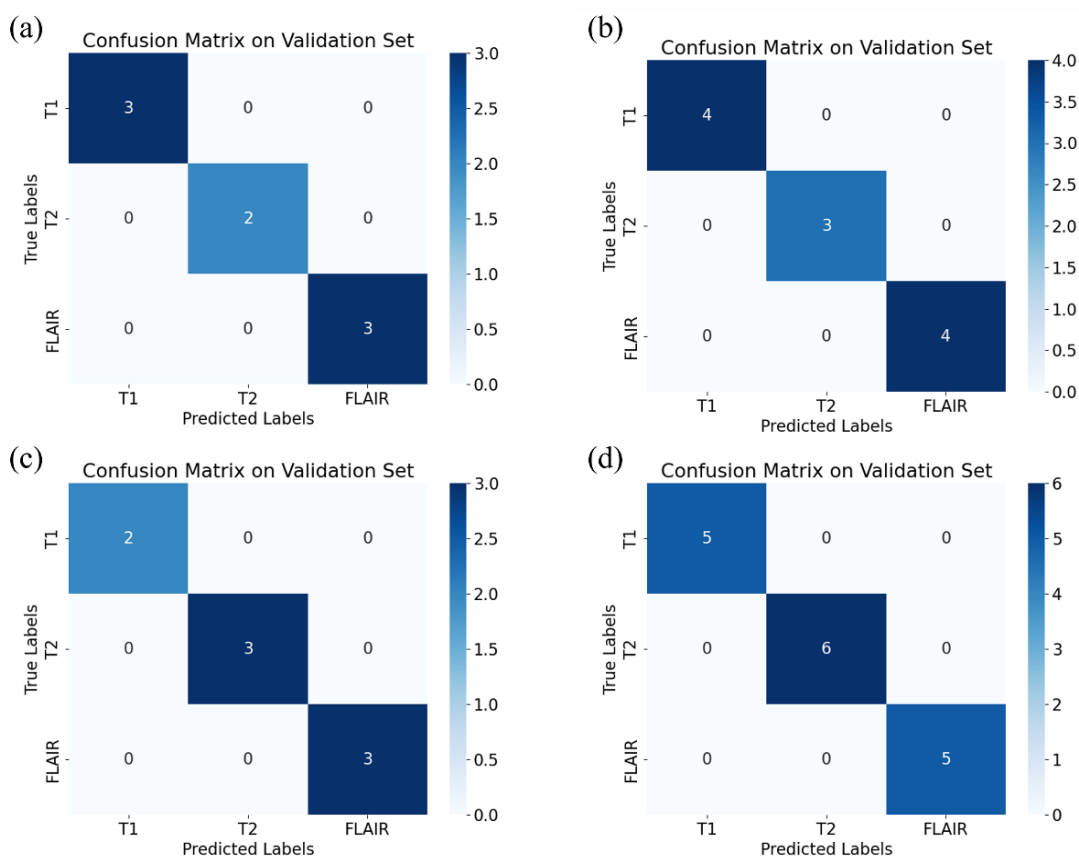

Figure 5-1: Comparison of confusion matrices for models trained on different datasets in the validation set. (a) Confusion matrix for the model trained on 0.35T real data; (b) for the model trained on 3T real data; (c) for the model trained on synthetic data; (d) for the model trained on 0.35T real data + synthetic data. The extremely high accuracies observed here highlight the potential risk of overfitting with scarce training data, and independent test set evaluation is needed.

## 5.4.2 Test Results

After training, the saved model weights are loaded. Four evaluation metrics—accuracy, precision, recall, and F1 score—are used to test the models on the test set. As commonly used classification metrics, accuracy reflects the proportion of correct predictions across all samples; precision calculates the probability that a predicted positive is truly positive; recall measures the proportion of real positives correctly identified; and F1 is the harmonic mean of precision and recall, comprehensively evaluating model performance. All indicators range from 0 to 1. The higher the value, the greater the consistency between predicted and real labels, and the better the model's classification accuracy.

The test results are shown in Table 5-2. The first three models have similar accuracy around 70%. The model trained with 0.35T real data performs slightly better in precision; synthetic data training gives higher recall; the F1 scores of the first and third models are close. The fourth model, which augments 0.35T data with synthetic low-field images from the fine-tuned model, achieves an accuracy of 96.15%, indicating that the generated images in this study can improve classification model accuracy, with precision, recall, and F1 also being the best. Also, the test set and the first model's training data come from the same dataset, giving the first model an accuracy of 71.15%, while the third model—trained with generated images—achieves 69.87%, showing that when training data is insufficient or unavailable, SD model-generated images can be used for data augmentation and as virtual data for model training.

Table 5-2 Performance of Trained Models on the Test Set

| Model | Accuracy | Precision | Recall | F1 Score |
|---|---|---|---|---|
| 0.35T Real Data Training Model | 0.7115 | 0.8325 | 0.6087 | 0.7030 |
| 3T Real Data Training Model | 0.6795 | 0.7015 | 0.6170 | 0.6566 |
| Synthetic Data Training Model | 0.6987 | 0.7001 | 0.6903 | 0.6952 |
| 0.35T Real Data + Synthetic Data Training Model | 0.9615 | 0.9630 | 0.9591 | 0.9614 |

Figure 5-2 compares confusion matrices for models trained on different datasets. (a) shows the confusion matrix for the model trained on 0.35T real data, where over half the T1 images are misclassified as FLAIR and two as T2; (b) is for the model trained on 3T real data, with most T1 and T2 classified correctly, but only moderate results for FLAIR, as it is often predicted as T2; (c) is for the model trained on synthetic data, with T1 and T2 faring slightly worse than 3T, but improved FLAIR classification; (d) is for the model trained with a mix of 0.35T real and synthetic data, achieving the best performance—most images from all three modalities are well distinguished, with three T1 misclassified as T2, one as FLAIR, two T2 as FLAIR, and all FLAIR classified correctly. This clearly demonstrates the superior classification capacity of the fourth model.

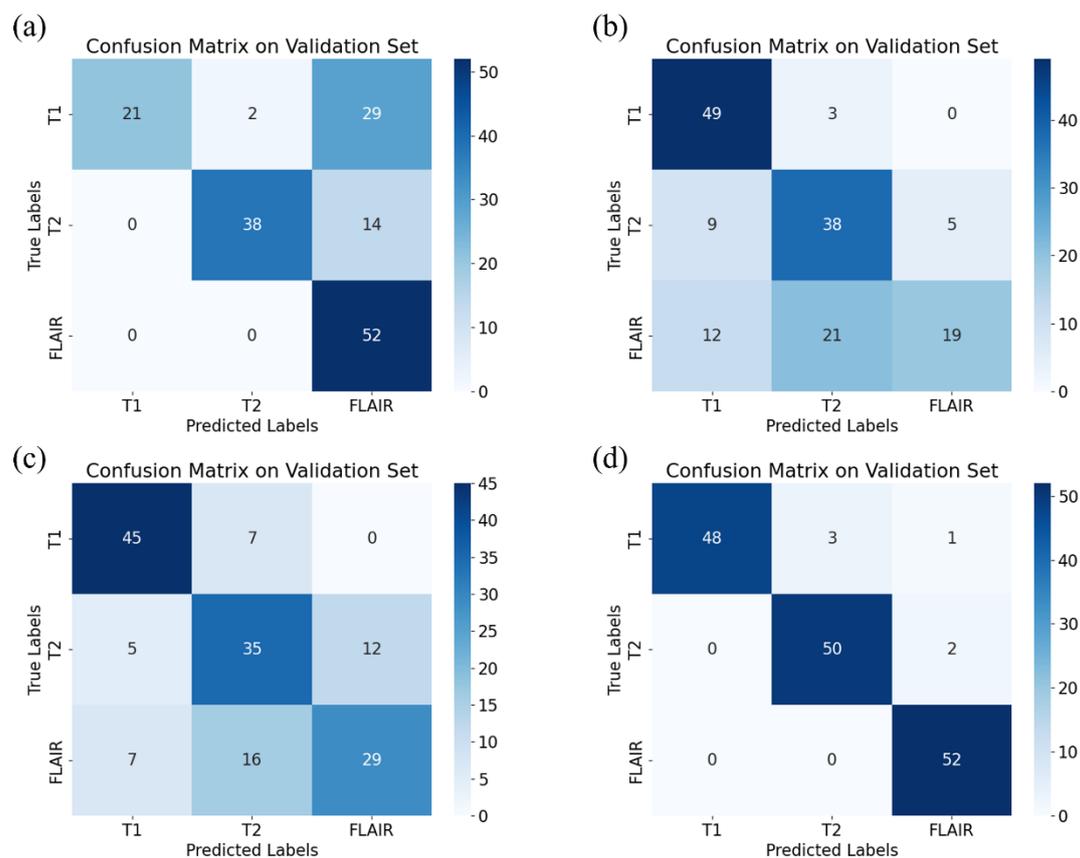

Figure 5-2 Comparison of confusion matrices of models trained on different training sets on the test set. (a) Confusion matrix of the model trained with 0.35T real data; (b) Confusion matrix of the model trained with 3T real data; (c) Confusion matrix of the model trained with synthetic data; (d) Confusion matrix of the model trained with 0.35T real data + synthetic data.

From the experimental data, the effect of using only simulated data is not particularly good. Although both the test set and the SD model-generated training set are low-field-strength MRI images, the two datasets are from different brands, field strengths, and locations. The training set of the SD model was collected using a Xingaoyi brand 0.3T scanner in Shenzhen, while the test set was collected using an Anke brand 0.35T scanner in Shaanxi, with some differences in specific imaging parameters. In addition, the training set samples of the SD model come from healthy young people, whereas the test set samples are from elderly patients. However, these synthetic data generally possess the typical features of low-field MRI images and can serve as a supplement to real data to improve the training effectiveness of classification models.

According to the data in Table 5-2, the best model performance is obtained when 0.35T data and synthetic data are used together for training. To compare the differences between this model and other models, samples that are well predicted by this model but poorly predicted by others are selected for display. As shown in Table 5-3, the first three rows present images and labels that are only correctly predicted by the model trained on 0.35T data and synthetic data, while all other models predict incorrectly; the last three rows show images and labels that are misclassified by some models. The displayed images include T1, T2, and FLAIR modalities. The model trained on both 0.35T data and synthetic data can correctly classify samples when other models make errors, which further demonstrates that adding synthetic data to the original training data can improve image classification accuracy.

Table 5-3 Comparative display of some classification results on different models

| Image | True Label | Predicted Labels | | | |
|---|---|---|---|---|---|
| | | Model trained with 0.35T data | Model trained with 3T data | Synthetic Data Training Model | Model trained with 0.35T data + synthetic data |
| 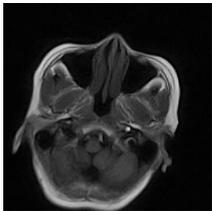 | T1 | T2 | T2 | T2 | T1 |
| 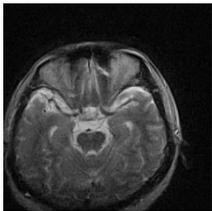 | T2 | FLAIR | FLAIR | FLAIR | T2 |
| 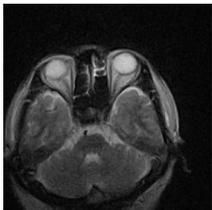 | T2 | FLAIR | FLAIR | FLAIR | T2 |

| Image | True Label | Predicted Labels | | | |
|---|---|---|---|---|---|
| | | Model trained with 0.35T data | Model trained with 3T data | Synthetic Data Training Model | Model trained with 0.35T data + synthetic data |
| 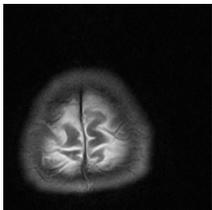 | T2 | T2 | T1 | T1 | T2 |
| 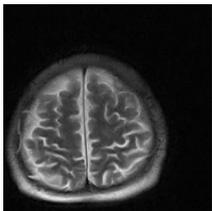 | T2 | T2 | T2 | T1 | T2 |
| 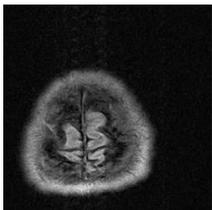 | FLAIR | FLAIR | T2 | T2 | FLAIR |

This experiment uses a small-sample image dataset for the classification task, demonstrating the value and potential of images generated by the SD model in practical applications. The classification data used in this experiment are different from the M4Raw and fastMRI datasets and are suitable for showcasing the value of the SD model. In addition, the main purpose of this experiment is to verify the feasibility of modality classification using SD model-generated images. Although the experimental results show good performance on small samples, to fully assess the model's generalization ability, further research on a larger scale dataset is needed in the future.

# 6. Conclusions and Prospects

## 6.1 Main Conclusions

Difficulties in acquiring MRI image training data limit related research. This study focuses on text-prompted MRI image generation. MRI technology provides multimodal anatomical and functional information and is an important technique in medical imaging, but data acquisition is hindered at the intersection of AI and medical imaging. To overcome such problems, this study explores MRI image generation algorithms based on the Stable Diffusion model, introduces text-conditioned mechanisms to establish semantic associations between text descriptions and MRI images, fine-tunes pretrained models, and achieves controllable generation of multi-field-strength, multi-modal MRI features from textual descriptions. The algorithm is based on a latent diffusion model that encodes images into low-dimensional latent spaces and denoises iteratively, reducing the computational complexity of high-resolution medical image generation. Experiments show that the algorithm can generate diverse MRI images similar to real images, with evaluation indicators significantly improved over the non-fine-tuned model. Good generative performance is achieved under different modalities and field strengths, reflecting a strong generalization ability. Finally, to demonstrate the model's value in practical application, experiments on different modality MRI image classification were conducted. The results indicate that images generated by the fine-tuned model can improve classification accuracy.

## 6.2 Areas for Improvement in This Study

Although this study has achieved results in text-prompted MRI image generation, there is still room for improvement to further enhance algorithm performance and application value. In data set selection, in addition to current healthy brain MRI images, images of different tumor types and other abnormalities can be added to further test the model's generalization and generative ability. In data preprocessing, central cropping is used to standardize images to 256×256, which inevitably leads to the loss of some image information, especially at the edges—as seen clearly in T1 modality and FLAIR images with 3T scanners. Future research could adopt scaling operations to unify image size and retain more information. Regarding model training strategy, this study only fine-tuned UNet and did not fine-tune the text encoder. In the future, joint fine-tuning of UNet and the text encoder can be attempted to enhance the semantic consistency of generated images. The general-purpose pretrained text encoder used in this study can also be replaced with domain-specific text encoders that are more capable in medical text understanding, such as RadBERT [38], and compared to the general text encoder.

## 6.3 Outlook

Text-prompted MRI image generation algorithms have broad application prospects and research potential in the field of medical imaging. In this study, by introducing a text-conditioning mechanism and the Stable Diffusion model, MRI images are generated from text. However, this research is still in an exploratory stage, and there is room for improvement in image detail fidelity, semantic consistency, and clinical applicability. Its potential application areas and future directions merit in-depth study.

Augmenting training datasets is one of the important research directions. Although relevant studies have made progress in the generation and application of synthetic data, there is still great potential for expansion in this field. Generated images can both augment the training dataset and protect patient privacy, and provide diverse training samples for models. Recent research in this direction includes work by Professor Zhang Kang's team at Wenzhou Medical University and Professor Qu Jia's team together with teams from Peking University, Macau University of Science and Technology, and others. Using their independently developed MINIM model [39], they incorporated generated image data into the original dataset and found that it significantly improved the classification accuracy of multiclass diagnostic models. This shows that generated images have important value in medical imaging. Future research can further explore how to optimize generated image data to better match the feature distribution of real data and improve model performance in practical applications.

Beyond data augmentation, the SD model can also play a key role in tasks such as MRI image reconstruction and denoising. For example, it can be used for image reconstruction in fast MRI scanning. By leveraging the generative capabilities of the SD model, high-quality MRI images can be reconstructed from undersampled k-space data [25]. In addition, SD models can be used in medical education to generate representative MRI images under different imaging conditions. These generated images can serve as teaching resources to help students better understand MRI principles and applications, rather than being limited to textbook and clinical case images. MRI image generation algorithms can also be integrated into MRI simulation scanning software to flexibly generate realistic MRI images based on different user settings.